\documentclass[journal, 10pt]{IEEEtran}
\IEEEoverridecommandlockouts
\usepackage{cite}
\usepackage{amsmath,amssymb,amsfonts}
\usepackage{graphicx}
\usepackage[caption=false,font=footnotesize]{subfig}
\usepackage{textcomp}
\usepackage{xcolor}
\usepackage{booktabs}
\usepackage{tabularx}
\usepackage{multirow}
\usepackage{diagbox}
\usepackage[binary-units]{siunitx} 
\DeclareSIUnit{\bps}{bps}
\usepackage{amsthm}
\usepackage{comment}

\def\tcr{\textcolor{red}}
\def\tr{\mathrm{tr}}
\newtheorem{proposition}{Proposition}
\usepackage[ruled,vlined]{algorithm2e}

\begin{document}

\title{Hierarchical Cell-Free Massive MIMO:\\A Simplified Design for Uniform Service Quality}
\author{Wei~Jiang,~\IEEEmembership{Senior~Member,~IEEE}
        and Hans~Dieter~Schotten
\thanks{Part of this work was presented at the 2024 IEEE International Conference on Communication (ICC-2024) \cite{Ref_jiang2024hierarchical}. (\textit{Corresponding author: Wei Jiang}) %The associated editor coordinating the review of this paper and approving it for publication was xxxxxx. (\textit{Corresponding author: Wei Jiang})
}
\thanks{W. Jiang and H. D. Schotten are with the German Research Center for Artificial Intelligence (DFKI), Kaiserslautern, Germany, and are also with University of Kaiserslautern (RPTU), Paul-Ehrlich Street, Kaiserslautern, 67663 Germany (e-mail: wei.jiang@dfki.de; schotten@eit.uni-kl.de).}
}

\maketitle

\begin{abstract}
In traditional cellular networks, users at the cell edge often suffer from poor quality of service (QoS) due to large distance-dependent path loss and severe inter-cell interference. While cell-free (CF) massive multi-input multi-out (MIMO) mitigates this issue by distributing access points (APs) to ensure uniform QoS, the deployment of numerous distributed APs and a fronthaul network incurs high infrastructure costs. To balance performance and cost efficiency, this article proposes a simplified design called hierarchical cell-free (HCF) massive MIMO. The key idea is to reduce the number of APs, thus minimizing the scale of the fronthaul network. The antennas from the decommissioned APs are aggregated at a central base station (cBS), which also serves as the coordinator for distributed APs. We derive closed-form expressions for uplink and downlink spectral efficiency (SE) for HCF, CF, and cellular massive MIMO under pilot contamination and correlated fading channels, considering the use of multi-antenna APs. Numerical results confirm that the hierarchical architecture achieves $95\%$-likely per-user SE comparable to CF, enhancing cell-edge user rates in cellular systems by over 100 times, while significantly reducing the complexity and cost of the fronthaul network in CF. We develop max-min fairness algorithms for joint power control of the cBS and APs in the downlink, and the users in the uplink. These algorithms not only boost fairness and system capacity but also dramatically lower transmission power, e.g., achieving over $70\%$ savings in uplink, particularly beneficial for battery-powered mobile devices.
\end{abstract}

\IEEEpeerreviewmaketitle

\section{Introduction}

In traditional cellular networks, each cell is served by a base station (BS) positioned at its center. Users close to the BS generally experience high quality of service (QoS). Conversely, users at the cell edge often encounter poor QoS due to factors such as large signal attenuation over distance, severe interference from neighboring cells, and handover inefficiencies inherent to cellular systems \cite{Ref_jiang2024TextBook}. As outlined in the ITU-R M.2410 recommendation \cite{Ref_non2017minimum}, the peak spectral efficiency (SE) targets for fifth-generation (5G),  typically achieved in the cell center, are set at \SI{30}{\bps\per\hertz{}} in the downlink and \SI{15}{\bps\per\hertz{}} in the uplink.  In contrast, the $95\%$-likely SE \cite{Ref_nayebi2017precoding} — also known as the $5^{th}$ percentile user rate \cite{Ref_ngo2017cellfree} or $5\%$-outage rate — serves as a key metric for evaluating cell-edge performance \cite{Ref_jiang2023celledge}. 5G's targets for this metric are notably lower, set at \SI{0.3}{\bps\per\hertz{}} (downlink) and \SI{0.21}{\bps\per\hertz{}} (uplink) in indoor hotspots, further reduced to \SI{0.12}{\bps\per\hertz{}} and \SI{0.0453}{\bps\per\hertz{}}, respectively, in rural areas. This stark contrast reveals a substantial performance gap, over a 100-fold difference, between users at the cell center and edge, underscoring the critical need for improved cell-edge QoS in next-generation networks  \cite{Ref_jiang2021road}.

Cell-free (CF) massive multi-input multi-output (MIMO) has arisen as a groundbreaking approach in cellular network design, attracting much attention from both academia and industry \cite{Ref_jiang20236GCH9}. Departing from conventional cell-based structures, CF deploys a large number of distributed access points (APs) spread throughout the coverage area. These APs, coordinated by a central processing unit (CPU) via a fronthaul network, work collaboratively to provide uniform QoS to all users. 
Many advancements in CF massive MIMO technology have been achieved in recent years. Early work by Ngo et al. \cite{Ref_ngo2017cellfree} validated its ability to effectively mitigate the cell-edge problem and ensure uniform QoS by benchmarking the $95\%$-likely user rate against small-cell systems. Concurrently, Nayebi et al. \cite{Ref_nayebi2017precoding} evaluated downlink performance under different linear precoding and power allocation algorithms. To address scalability challenges, Björnson et al. \cite{Ref_bjornson2020scalable} systematically reviewed dynamic cooperation clustering and network densification limits. Interdonato et al. \cite{Ref_interdonato2019downlink} later demonstrated that the use of downlink pilots can improve performance despite introducing overhead and exacerbating pilot contamination. In \cite{Ref_ngo2018total}, Ngo et al. developed an optimization method to maximize total energy efficiency under per-user SE and per-AP power constraints. Björnson and Sanguinetti \cite{Ref_bjornson2020making} provided a comprehensive  analysis under four levels of uplink cooperation, revealing the advantage of centralized minimum mean-square error (MMSE) processing.

While CF offer the potential for improved performance and fairness, the deployment of numerous distributed APs and a fronthaul network incurs high infrastructure costs. Traditional wireless networks already face high deployment costs and challenges related to site acquisition. CF systems exacerbate these issues by requiring hundreds of wireless AP sites in a small area \cite{Ref_masoumi2020performance}. Additionally, installing and operating a large-scale fronthaul network, typically based on fiber-optic cables, to connect a massive number of APs leads to substantial capital and operational costs and may face deployment restrictions in certain areas. Recent research has proposed several strategies to overcome this obstacle, with key approaches focusing on the fronthaul network. For instance, Furtado et al. \cite{furtado2022cell} discussed the challenges of a non-scalable complex and costly fronthaul network and identified cost-saving fronthaul topologies. Buzzi et al. proposed user-centric approaches \cite{Ref_buzzi2017cellfree, Ref_buzzi2020usercentric} that reduce fronthaul overhead by dynamically clustering APs around users, maintaining performance with reduced complexity. Infrastructure costs can also be optimized through AP design: Jiang and Schotten \cite{Ref_jiang2024cost} demonstrated that equipping APs with multiple antennas reduces the total number of required wireless AP sites, thereby lowering fronthaul costs. 

Wireless fronthaul alternatives have emerged as viable solutions to eliminate the high costs associated with infrastructure construction, such as trenching, fiber cable laying, and life-cycle maintenance. Demirhan and Alkhateeb \cite{demirhan2022enabling} proposed to replace optical fiber fronthaul with millimeter-wave wireless fronthaul to minimize infrastructure costs and deployment time. Similarly, Jazi et al. \cite{jazi2023integrated} explored integrated access and backhaul via spectrum sharing, optimizing fronthaul efficiency in dense networks. For industrial settings, Li et al. \cite{li2023two} introduced a two-tier terahertz fronthaul architecture, eliminating wired connections in indoor CF deployments. The challenges under constrained fronthaul capacity have also driven innovations in signal processing. For instance,  Kim et al. \cite{kim2022cellfree} proposed a precoding scheme with max-min power allocation using fronthaul compression and low-resolution digital-to-analog converters. Parida and Dhillon \cite{parida2022cell} quantified the impact of finite fronthaul capacity on downlink performance, motivating advanced quantization techniques such as scalable multivariate compression \cite{park2024scalable}. Ranjbar et al. \cite{ranjbar2024cell} further analyzed sequential topologies (e.g., daisy chains) under hardware limitations, emphasizing trade-offs between memory constraints and optimal design. Resource allocation strategies, such as meta-heuristic bit allocation frameworks \cite{kim2023meta}, have also been developed to maximize fronthaul efficiency under strict capacity limits. 

Wireless backhauling is by far the most cost-effective solution, but its adoption does not come for free \cite{siddique2017downlink}. \textit{Out-of-band} implementations require each AP to be equipped with an additional transceiver and antenna or antenna array. Also, it potentially incurs regulatory licensing fees for the specific frequency band. Conversely, \textit{in-band} frequency usage requires the access link to allocate approximately half of its frequency/time resources to the fronthaul link, significantly degrading system performance. To optimize the cost-effectiveness, this article proposes a simplified design called hierarchical cell-free (HCF) massive MIMO. Our design deliberately reduces the number of APs so as to minimize the scale of the fronthaul network. The antennas from the decommissioned APs are aggregated to a centrally located BS, which we refer to as the central base station (cBS) to distinguish it from the traditional paradigm. As the antennas and transceivers are reused from the retired APs, HCF avoids additional hardware costs.  The cBS not only serves as a central signal transceiver but also replaces the CPU in CF systems to coordinate the distributed APs.

The main contributions of this work can be listed as follows:
\begin{itemize}
    \item This article is the first to introduce a hierarchical pattern that smoothly bridges the gap between fully centralized (cellular) and fully distributed (CF) massive MIMO. Through limiting the number of distributed APs, it effectively lowers the scale of the fronthaul network, thus reducing complexity and cost. cBS is introduced to retain the centralization gain without incurring deployment costs, as it repurposes hardware from the decommissioned APs. 
    \item We develop HCF max-min fairness algorithms for joint power control of the cBS and APs in the downlink, and the users in the uplink. These power controls not only enhance fairness and system capacity but also substantially save the transmission power. 
    \item We derive closed-form SE expressions for not only HCF, but also CF and cellular, in both downlink and uplink operations. Our analysis accounts for multi-antenna APs, spatially correlated channels, and pilot contamination. It extends beyond prior CF works, which were mostly constrained to single-antenna APs or independent channels.        
    \item Comprehensive numerical evaluation is conducted to compare HCF, CF, and cellular massive MIMO in both uplink and downlink,  considering key performance metrics including $95\%$-likely per-user SE (fairness), sum throughput (capacity), and power saving (energy efficiency), while accounting for various factors, such as micro vs. macro cells, pilot contamination, angular spread, and equal/max-min power optimization.
\end{itemize}

The paper is organized as follows: Section II models the HCF system. Section III discusses uplink training in the context of channel correlation and pilot contamination. Section IV analyzes uplink data transmission, deriving closed-form SE expressions and proposing max-min power optimization. Section V extends this analysis to downlink transmission. Section VI details the simulation setups and presents key numerical results. Finally, Section VII offers conclusions.

\textbf{Notations}: Throughout this article, \textbf{boldface} lowercase and uppercase letters denote vectors and matrices, respectively. The following operators are used for their operations: 
$(\cdot)^*$ (conjugate), $(\cdot)^T$ (transpose), $(\cdot)^H$ (Hermitian transpose), $(\cdot)^{-1}$ (inverse), and $\tr(\cdot)$ (the trace of a matrix).  The Frobenius norm is represented by $\|\cdot\|$, $\log_{2}$ specifies a logarithm with base 2, and $\lg$ is the common logarithm with base 10. Statistical expectation is denoted by $\mathbb{E}$, and $\mathbf{I}$ refers to an identity matrix of suitable dimensions. The symbol $\in$ indicates set membership, while $ \setminus $ represents the set difference, e.g., $k'\in \mathbb{K} \setminus \{k\}$  means  $k'$ is an element of the set $\mathbb{K}$ excluding the specific element $k$. Estimated variables are marked with a hat (e.g., $\hat{x}$), and estimation errors are distinguished by a tilde (e.g., $\tilde{x}$). The notation $\mathcal{CN}$ corresponds to a complex Gaussian distribution.

\section{System Model}

\begin{figure}[!t]
    \centering 
    \subfloat[CF: a large number of distributed APs cooperatively cover an area, coordinated by a CPU through a large-scale fronthaul network.]{ \label{fig:CF}
    \includegraphics[width=0.45\textwidth]{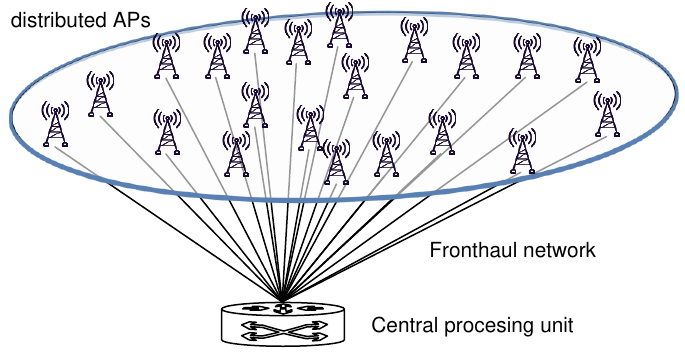} 
    }    \\
    \subfloat[HCF: a cBS equipped with a massive antenna array, aided by some distributed eAPs, which are connected to the cBS. ]{ \label{fig:heteromimo}
    \includegraphics[width=0.435\textwidth]{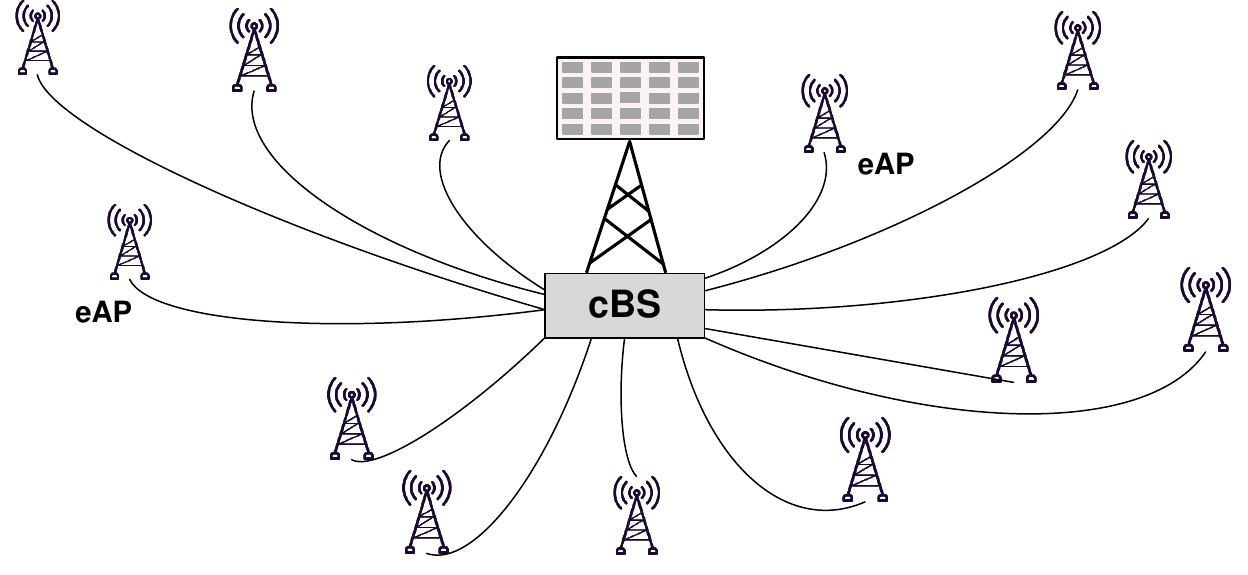} 
    }
    \caption{A comparative illustration of cell-free and hierarchical cell-free massive MIMO networks.  }
    \label{fig:SystemModel}
\end{figure}

In CF massive MIMO, a network of $M$ massive antennas is distributed across a coverage area to serve $K$ users, as depicted in \figurename \ref{fig:CF}. To exploit the advantages of channel hardening and favorable propagation \cite{Ref_marzetta2010noncooperative}, the number of service antennas must substantially exceed the number of users, i.e., $M\gg K$. User equipment (UE) is typically equipped with a single antenna, while each AP may have either a single or multiple antennas. A CPU manages all APs via a fronthaul network. In this paper, we consider a hierarchical design for CF massive MIMO, as shown in Fig.~\ref{fig:heteromimo}. The proposed  architecture features a cBS with $N_{b}$ co-located antennas, positioned near the center of the coverage area. The network also includes $L$ distributed APs, each equipped with $N_a$ antennas. To distinguish from the conventional paradigm, these are referred to as edge access points (eAPs) hereinafter. Unlike conventional base stations, the cBS also functions as the CPU for coordinating the eAPs.  We assume that HCF maintains the same total number of antennas as CF, such that $N_b + L N_a = M$. The aim of this design is to simplify the fronthaul network, thus lowering complexity and cost,  as only a subset of the service antennas require connections.

The index sets of eAPs and users are denoted by $\mathbb{L}= \{1,\ldots,L\}$ and $\mathbb{K}=\{1,\ldots,K\}$, respectively.
The channel between eAP $l$, $\forall l\in \mathbb{L}$ and user $k$, $\forall k \in \mathbb{K}$ is denoted by $\mathbf{h}_{kl}\in \mathbb{C}^{N_a}$. 
Under the standard block fading model, each \textit{coherent block} is a time-frequency interval of $\tau_c$ channel uses during which the channel response remains approximately constant. Each coherence block applies an independent realization from a \textit{correlated}\footnote{Most prior works on CF have simplified analyses by assuming either single-antenna APs or multi-antenna APs with independent channels among an AP's co-located antennas. However, in practical scenarios, closely spaced antennas at a multi-antenna AP exhibit correlated channel responses. Thus,  it is essential to consider channel correlations for precious performance evaluation and system design.} Rayleigh fading distribution $\mathbf{h}_{kl} \sim \mathcal{CN}(\mathbf{0}, \mathbf{R}_{kl} )$. $\mathbf{R}_{kl}$ is the spatial correlation matrix that characterizes the channel's spatial properties, defined as $\mathbf{R}_{kl}=\mathbb{E}[  \mathbf{h}_{kl} \mathbf{h}_{kl}^H ]$.  Large-scale fading coefficient, $\beta_{kl} = \tr(\mathbf{R}_{kl})/N_{a}$, accounts for geometric path loss and shadowing. The generation of the channel realization is based on $\mathbf{h}_{kl}=\mathbf{R}_{kl}^{1/2}\mathbf{h}_{kl}'$, where $\mathbf{h}_{kl}'$ denotes independent and identically distributed Rayleigh fading with zero-mean, unit-variance elements \cite{yu2004modeling}.
Similarly, we write $\mathbf{h}_{k0}\in  \mathbb{C}^{N_b}$ to denote the channel between the cBS and user $k$, $\forall k \in \mathbb{K}$. It follows the distribution $ \mathcal{CN}(\mathbf{0}, \mathbf{R}_{k0})$, where $\mathbf{R}_{k0}$ is the spatial correlation matrix, and its corresponding large-scale fading coefficient equals $\beta_{k0}=\tr(\mathbf{R}_{k0})/N_{b}$. Since spatial correlation matrix does not change rapidly over time or frequency, the system can measure it on a long-term basis and distributes it periodically. Thus, it is reasonable to assume that all nodes have perfect knowledge of this information.

To avoid the prohibitive overhead of downlink pilots, which scales with the number of service antennas, time-division duplexing (TDD) is employed in massive MIMO to separate the downlink and uplink transmission. Thus, each coherent block is split into three phases: uplink training, uplink data transmission, and downlink data transmission. In the downlink, all eAPs and cBS transmit data symbols over the same time-frequency resource, while all UEs simultaneously send their signals in the uplink at another instant.

\section{Uplink Training} \label{section_ULtraining}

Assume pilot sequences span $\tau_p$ symbols, allowing for a maximum of $\tau_p$ orthogonal sequences, which are collectively represented as  $\boldsymbol \Omega = [\boldsymbol \omega_{1}, \ldots, \boldsymbol \omega_{\tau_p}] \in \mathbb{C}^{\tau_p \times \tau_p}$. These sequences have unit-magnitude elements for a constant power level, implying that $\boldsymbol \Omega \boldsymbol \Omega^H = \tau_p \mathbf{I}_{\tau_p}$. 
During the uplink training phase, each UE transmits a pilot sequence. Denoting the pilot sequence of UE $k$ by $\mathbf{i}_k$, we have $\exists ! x{\in}\{1,\ldots,\tau_p\}\: \left( \mathbf{i}_k = \boldsymbol \omega_{x}\right)$, where $\exists !$ stands for unique existential quantification.

\subsubsection{eAP-UE Channel Estimation}
When the UEs simultaneously transmit their pilot sequences,  the received pilot signal at a typical eAP $l$ is given by
\begin{equation} 
    \boldsymbol{\Psi}_l = \sqrt{p_u} \sum\nolimits_{k\in \mathbb{K}} \mathbf{h}_{kl} \mathbf{i}_k^T+\mathbf{Z}_{l},
\end{equation}
where $p_u$ is UE's transmit power constraint and $\mathbf{Z}_l\in \mathbb{C}^{N_a \times \tau_p}$ represents additive noise with independent entries following the distribution of $\mathcal{CN}(0,\sigma_z^2)$, where $\sigma_z^2$ is the noise power. 
From the scalability perspective, some UEs need to share the same sequence if $\tau_p<K$, when the system simultaneously serves a lot of active users. This leads to pilot contamination \cite{elijah2015comprehensive}. We use $\mathcal{P}_k$ to denote the set of indices for the users, including user $k$ itself, that utilize the same sequence of $\mathbf{i}_k$. 

To estimate $\mathbf{h}_{kl}$, correlate $\boldsymbol{\Psi}_l$ with the known pilot sequence $\mathbf{i}_k$, resulting in\footnote{In this paper, a \textit{general} user is denoted by $k'$, while a \textit{particular} user, which is under analysis and discussion, is represented by $k$. }
\begin{align} \label{EQN_channelestimate} \nonumber
\boldsymbol{\psi}_{kl} & =\boldsymbol{\Psi}_l \mathbf{i}_k^* =  \sqrt{p_u}  \sum\nolimits_{k'\in \mathbb{K}} \mathbf{h}_{k'l} \mathbf{i}_{k'}^T \mathbf{i}_{k}^*+ \mathbf{Z}_{l}\mathbf{i}_k^*\\
&=\sqrt{p_u}\tau_p \mathbf{h}_{kl} +  \sqrt{p_u}\tau_p \sum\nolimits_{k'\in \mathcal{P}_k \setminus \{k\}} \mathbf{h}_{k'l} + \mathbf{Z}_{l}\mathbf{i}_k^*,
\end{align}
as $\mathbf{i}_{k'}^T \mathbf{i}_{k}^*=0$, when $k'\notin \mathcal{P}_k$.  
Conducting channel estimation with linear MMSE \cite{SIG-093}, the estimate of $\mathbf{h}_{kl}$ is obtained by
\begin{equation} \label{eQ_hestimate}
    \hat{\mathbf{h}}_{kl} = \mathbb{E}\left[  \mathbf{h}_{kl} \boldsymbol{\psi}_{kl}^H \right] \left( \mathbb{E}\left[  \boldsymbol{\psi}_{kl} \boldsymbol{\psi}_{kl}^H \right] \right)^{-1}  \boldsymbol{\psi}_{kl}.
\end{equation}
In our case, 
\begin{align} \nonumber
\mathbb{E}\left[  \mathbf{h}_{kl}  \boldsymbol{\psi}_{kl}^H \right] & = \sqrt{p_u} \tau_p \mathbb{E}\left[  \mathbf{h}_{kl}  \mathbf{h}_{kl} ^H \right]\\ \nonumber
&+ \sqrt{p_u} \tau_p \sum\limits_{k'\in \mathcal{P}_k \setminus \{k\}}  \mathbb{E}\left[  \mathbf{h}_{kl}  \mathbf{h}_{k'l} ^H \right] + \mathbb{E}\left[\mathbf{h}_{kl}\mathbf{Z}_{l}\mathbf{i}_k^*  \right]
\\ 
&=\sqrt{p_u} \tau_p \mathbf{R}_{kl},
\end{align}
and
\begin{align}   \nonumber
  \mathbb{E}\left[  \boldsymbol{\psi}_{kl} \boldsymbol{\psi}_{kl}^H \right] &= p_u\tau_p^2 \sum_{k'\in \mathcal{P}_k} \mathbb{E}\left[  \mathbf{h}_{k'l}  \mathbf{h}_{k'l}^H \right]  +  \mathbb{E}\left[  \mathbf{Z}_{l}\mathbf{i}_k \mathbf{i}_k^H \mathbf{Z}_{l}^H  \right] \\ 
  &= \tau_p \left(p_u\tau_p\sum\nolimits_{k'\in \mathcal{P}_k } \mathbf{R}_{k'l} + \sigma_z^2\mathbf{I}_{N_a}\right).
\end{align}

Hence, \eqref{eQ_hestimate} can be rewritten as 
\begin{equation} \label{eQn_hhat}
    \hat{\mathbf{h}}_{kl} =  \sqrt{p_u}  \mathbf{R}_{kl} \boldsymbol{\Gamma}_{kl}^{-1} \boldsymbol{\psi}_{kl},
\end{equation}
where 
\begin{equation}
     \boldsymbol{\Gamma}_{kl}=p_u\tau_p\sum \nolimits_{k'\in \mathcal{P}_k } \mathbf{R}_{k'l} + \sigma_z^2\mathbf{I}_{N_a}.
\end{equation}
The correlation matrix of $\hat{\mathbf{h}}_{kl}$ is computed by
\begin{align} \nonumber \label{EQN_correlationMatrixCHest}
\mathbb{E}\left[ \hat{ \mathbf{h}}_{kl} \hat{\mathbf{h}}_{kl}^H \right]&=\mathbb{E}\left[  \mathbf{h}_{kl} \boldsymbol{\psi}_{kl}^H \right] \left( \mathbb{E}\left[  \boldsymbol{\psi}_{kl} \boldsymbol{\psi}_{kl}^H \right] \right)^{-1} \mathbb{E}\left[  \boldsymbol{\psi}_{kl} \mathbf{h}_{kl}^H \right]\\
&=p_u \tau_p\mathbf{R}_{kl} \boldsymbol{\Gamma}_{kl}^{-1}\mathbf{R}_{kl},
\end{align}
implying that $\hat{\mathbf{h}}_{kl}$ follows a complex Gaussian distribution:
\begin{equation}
    \hat{\mathbf{h}}_{kl} \sim \mathcal{CN}(\mathbf{0}, p_u \tau_p \mathbf{R}_{kl} \boldsymbol{\Gamma}_{kl}^{-1}\mathbf{R}_{kl} ).
\end{equation}
The estimation error, defined as $\tilde{\mathbf{h}}_{kl}=\mathbf{h}_{kl} - \hat{\mathbf{h}}_{kl}$, arises due to additive noise and pilot contamination, following $\mathcal{CN}(\mathbf{0}, \boldsymbol{\Theta }_{kl} )$, where its correlation matrix is given by
\begin{equation}
\boldsymbol{\Theta}_{kl}=\mathbb{E}\left[ \tilde{ \mathbf{h}}_{kl} \tilde{\mathbf{h}}_{kl}^H \right]=\mathbf{R}_{kl} - p_u \tau_p \mathbf{R}_{kl} \boldsymbol{\Gamma}_{kl}^{-1}\mathbf{R}_{kl}. 
\end{equation}

Based on the system implementation, channel estimation can be conducted distributedly at each eAP or centrally at the cBS. For eAP $l$, transmitting \(\boldsymbol{\Psi}_l\) over the fronthaul requires \(N_a \times \tau_p\) complex scalars per coherent block. It is more efficient than performing \textit{local} channel estimation and then sending the channel estimates \(\{\hat{\mathbf{h}}_{kl}\}_{k \in \mathbb{K}}\), which require \(N_a \times K\) complex scalars, as \(K > \tau_p\) in most scenarios. Hence, we assume the use of centralized estimation in our article.

\subsubsection{cBS-UE Channel Estimation}
Likewise, the cBS sees
\begin{equation}
    \boldsymbol{\Psi}_{0} = \sqrt{p_u} \sum\nolimits_{k\in \mathbb{K}} \mathbf{h}_{k0} \mathbf{i}_k^T+\mathbf{Z}_{0},
\end{equation}
where $\mathbf{Z}_0\in \mathbb{C}^{N_b \times \tau_p}$ represents the receiver noise at the cBS. Its channel estimation process mirrors that of the eAPs. Consequently, we obtain the MMSE estimate 
\begin{equation} \label{eQN_hhatdistribution}
    \hat{\mathbf{h}}_{k0} \sim \mathcal{CN}\left(\mathbf{0}, p_u \tau_p \mathbf{R}_{k0} \boldsymbol{\Gamma}_{k0}^{-1}\mathbf{R}_{k0} \right ),
\end{equation}
where $\boldsymbol{\Gamma}_{k0} =  p_u\tau_p\sum_{k'\in \mathcal{P}_k } \mathbf{R}_{k'0} + \sigma_z^2\mathbf{I}_{N_b}$.
The estimation error follows the distribution of $\tilde{\mathbf{h}}_{k0} \sim \mathcal{CN}(\mathbf{0}, \boldsymbol{\Theta}_{k0} )$ with
\begin{equation}
    \boldsymbol{\Theta}_{k0}=\mathbb{E}\left[ \tilde{ \mathbf{h}}_{k0} \tilde{\mathbf{h}}_{k0}^H \right]= \mathbf{R}_{k0} - p_u \tau_p \mathbf{R}_{k0} \boldsymbol{\Gamma}_{k0}^{-1}\mathbf{R}_{k0}.
\end{equation}

\section{Uplink Data Transmission}
In the uplink, all UEs simultaneously transmit their information-bearing symbols, where UE \( k \) transmits \( x_k \) with a power coefficient $0\leqslant \eta_k \leqslant 1$. These symbols are zero-mean, have unit variance, and are mutually uncorrelated. That is, the covariance matrix of $\mathbf{x}=[x_1,\ldots,x_K]^T$ satisfies $\mathbb{E}[\mathbf{x}\mathbf{x}^H]=\mathbf{I}_K$. 

\subsection{Signal Detection for HCF Massive MIMO}
During each channel use, eAP \( l \) receives
\begin{equation} \label{GS_uplink_RxsignalAP}
    \mathbf{y}_l = \sqrt{p_u} \sum\nolimits_{k\in \mathbb{K}} \sqrt{\eta_k} \mathbf{h}_{kl} x_k + \mathbf{z}_l,
\end{equation} 
where the receiver noise $\mathbf{z}_l\sim \mathcal{CN}(\mathbf{0},\sigma^2_z\mathbf{I}_{N_a})$.
Meanwhile, the cBS observes 
\begin{equation} \nonumber
    \mathbf{y}_0 = \sqrt{p_u} \sum\nolimits_{k\in \mathbb{K}} \sqrt{\eta_k} \mathbf{h}_{k0} x_k + \mathbf{z}_0
\end{equation} 
with $\mathbf{z}_0\sim \mathcal{CN}(\mathbf{0},\sigma^2_z\mathbf{I}_{N_b})$.

For ease of exposition, we employ the matched filtering (MF) detection method, also referred to as maximum-ratio combining \cite{Ref_ngo2017cellfree}. Additionally, we assume centralized processing \cite{Ref_bjornson2020making}, where eAP $l$ transmits $\mathbf{y}_l$ to the cBS via the fronthaul network. To recover $x_k$, the cBS correlates $\mathbf{y}_0$ with the MF combining vector that equals to $\hat{\mathbf{h}}_{k0}$, and multiplies $\mathbf{y}_l$ by  $\hat{\mathbf{h}}_{kl}^H$ for all $l\in \mathbb{L}$. This produces a soft estimate of $x_k$ as
\begin{align}  \nonumber \label{eQx_UL_rxSig}
    \hat{x}_k & = \frac{1}{\sqrt{p_u}} \left(\hat{\mathbf{h}}_{k0}^H\mathbf{y}_{0} + \sum\nolimits_{l\in \mathbb{L}}\hat{\mathbf{h}}_{kl}^H\mathbf{y}_l \right ) \\ \nonumber & = \hat{\mathbf{h}}_{k0}^H\left(  \sum\nolimits_{k'\in \mathbb{K}} \sqrt{\eta_{k'}} \mathbf{h}_{k'0} x_{k'}  + \frac{1}{\sqrt{p_u}}\mathbf{z}_0\right)\\ & + \sum\nolimits_{l\in \mathbb{L}}\hat{\mathbf{h}}_{kl}^H\left(  \sum\nolimits_{k'\in \mathbb{K}} \sqrt{\eta_{k'}} \mathbf{h}_{k'l} x_{k'} + \frac{1}{\sqrt{p_u}}\mathbf{z}_l \right) .
\end{align}
\begin{proposition}
The achievable SE for user $k$ in the uplink of HCF massive MIMO systems is expressed as
\begin{equation}
    R_k^{hcf} = \left(1-\frac{\tau_p}{\tau_u}\right)\mathbb{E} \Bigl[\log_2 \left(1+\gamma_k^{hcf} \right) \Bigr],
\end{equation}
where the pre-log factor $1-\tau_p/\tau_u$ accounts for the overhead of uplink pilots, and the instantaneous effective signal-to-interference-plus-noise ratio (SINR) is given by
\begin{equation} \label{eQN_sinrULcentralized}
    \gamma_k^{hcf} = 
     \frac{  \eta_k \left( \|\hat{\mathbf{h}}_{k0}\|^2+ \sum_{l\in \mathbb{L}} \| \hat{\mathbf{h}}_{kl} \|^2 \right)^2  }{ \left\{ \begin{aligned}    &  \sum\limits_{k'\in \mathbb{K}}  \eta_{k'} \left(\hat{\mathbf{h}}_{k0}^H\boldsymbol{\Theta}_{k'0}\hat{\mathbf{h}}_{k0} +\sum \nolimits_{l\in \mathbb{L}} \hat{\mathbf{h}}_{kl}^H\boldsymbol{\Theta}_{k'l}\hat{\mathbf{h}}_{kl}\right) \\
    &+ 
    \sum\limits_{k'\in \mathbb{K}\backslash \{k\}} \eta_{k'} \left| \hat{\mathbf{h}}_{k0}^H \hat{\mathbf{h}}_{k'0} + \sum\nolimits_{l\in \mathbb{L}}\hat{\mathbf{h}}_{kl}^H \hat{\mathbf{h}}_{k'l} \right|^2 
    \\
    &+\frac{\sigma_z^2}{p_u} \left ( \|  \hat{\mathbf{h}}_{k0}\|^2 + \sum\nolimits_{l\in \mathbb{L}} \| \hat{\mathbf{h}}_{kl}\|^2 \right)  \end{aligned} \right\} }.
\end{equation}
\end{proposition}
\begin{IEEEproof}
To facilitate the derivation, \eqref{eQx_UL_rxSig} is rewritten as 
\begin{align}  \nonumber \label{eQx_UL_rxSig22222}
    \hat{x}_k  =&\underbrace{ \sqrt{ \eta_k} \left[\hat{\mathbf{h}}_{k0}^H \hat{\mathbf{h}}_{k0}+ \sum\nolimits_{l\in \mathbb{L}}\hat{\mathbf{h}}_{kl}^H \hat{\mathbf{h}}_{kl}\right] x_k }_{\mathcal{S}_0:\:desired\:signal} \\ \nonumber &+ \underbrace{  \sum\nolimits_{k'\in \mathbb{K}} \sqrt{\eta_{k'}} \left[\hat{\mathbf{h}}_{k0}^H \tilde{\mathbf{h}}_{k'0}+ \sum\nolimits_{l\in \mathbb{L}}\hat{\mathbf{h}}_{kl}^H \tilde{\mathbf{h}}_{k'l}\right]  x_{k'} }_{\mathcal{I}_1:\:channel\:estimation\:error}\\ \nonumber &+\underbrace{\sum\nolimits_{k'\in \mathbb{K}\backslash \{k\}} \sqrt{\eta_{k'}} \left[ \hat{\mathbf{h}}_{k0}^H \hat{\mathbf{h}}_{k'0} + \sum\nolimits_{l\in \mathbb{L}}\hat{\mathbf{h}}_{kl}^H \hat{\mathbf{h}}_{k'l} \right]  x_{k'}}_{\mathcal{I}_2:\:inter-user\:interference} \\ &+\underbrace{ \frac{1}{\sqrt{p_u}}\left(\hat{\mathbf{h}}_{k0}^H\mathbf{z}_0 + \sum\nolimits_{l\in \mathbb{L}}\hat{\mathbf{h}}_{kl}^H\mathbf{z}_l \right ) }_{\mathcal{I}_3:\:colored\:noise}.
\end{align}
Due to the independence among data symbols, channel estimates, and estimation errors, the interference terms $\mathcal{I}_1$, $\mathcal{I}_2$, and $\mathcal{I}_3$ exhibit mutual uncorrelation. As stated in \cite{Ref_hassibi2003howmuch}, the worst-case noise for mutual information corresponds to Gaussian additive noise with a variance equal to the sum of the variances of $\mathcal{I}_1$, $\mathcal{I}_2$, and $\mathcal{I}_3$. Thus, the effective SINR is given by
\begin{equation}  \label{cfmmimo:formularSNR} 
    \gamma_k^{hcf}  {=} \frac{|\mathcal{S}_0|^2}{\mathbb{E}\left[|\mathcal{I}_1+\mathcal{I}_2+\mathcal{I}_3|^2\right]} {=} \frac{|\mathcal{S}_0|^2}{\mathbb{E}\left[|\mathcal{I}_1|^2\right]+\mathbb{E}\left[|\mathcal{I}_2|^2\right]+\mathbb{E}\left[|\mathcal{I}_3|^2\right]},
\end{equation} 
where the power gain of the desired signal equals
\begin{align} \nonumber
    |\mathcal{S}_0|^2  &=  \eta_k  \left |\hat{\mathbf{h}}_{k0}^H \hat{\mathbf{h}}_{k0}+ \sum\nolimits_{l\in \mathbb{L}}\hat{\mathbf{h}}_{kl}^H \hat{\mathbf{h}}_{kl}\right |^2 \\
    &=  \eta_k  \left ( \|\hat{\mathbf{h}}_{k0}\|^2+ \sum\nolimits_{l\in \mathbb{L}} \| \hat{\mathbf{h}}_{kl} \|^2\right )^2.
\end{align}

Since $\{\hat{\mathbf{h}}_{k0}\}_{k\in \mathbb{K}}$ and $\{\hat{\mathbf{h}}_{kl}\}_{k\in \mathbb{K}, l\in \mathbb{L}}$ are \textit{deterministic} for the cBS, see \cite{SIG-093}, the conditional variance of $\mathcal{I}_1$ is  derived as
\begin{align}  \label{eQN_deriv0001}
     \mathbb{E}\left[|\mathcal{I}_1|^2\right]  & \overset{(a)}{=}  \sum_{k'\in \mathbb{K}} \eta_{k'} \mathbb{E}\left[\left| \hat{\mathbf{h}}_{k0}^H \tilde{\mathbf{h}}_{k'0}+ \sum\nolimits_{l\in \mathbb{L}}\hat{\mathbf{h}}_{kl}^H \tilde{\mathbf{h}}_{k'l}   \right|^2 \right] \\ \nonumber
     & \overset{(b)}{=}  \sum_{k'\in \mathbb{K}} \eta_{k'} \left( \mathbb{E}[| \hat{\mathbf{h}}_{k0}^H \tilde{\mathbf{h}}_{k'0}|^2 ] + \sum_{l\in \mathbb{L}} \mathbb{E}[|\hat{\mathbf{h}}_{kl}^H \tilde{\mathbf{h}}_{k'l}   |^2 ] \right)  \\ \nonumber
     & \overset{(c)}{=}  \sum\nolimits_{k'\in \mathbb{K}} \eta_{k'}  \left(  \begin{aligned}       
    &\hat{\mathbf{h}}_{k0}^H\mathbb{E}\left[  \tilde{\mathbf{h}}_{k'0} \tilde{\mathbf{h}}_{k'0}^H  \right]\hat{\mathbf{h}}_{k0} \\
    &+ \sum\nolimits_{l\in \mathbb{L}} \hat{\mathbf{h}}_{kl}^H \mathbb{E}\left[ \tilde{\mathbf{h}}_{k'l} \tilde{\mathbf{h}}_{k'l}^H   \right]  \hat{\mathbf{h}}_{kl} 
     \end{aligned} \right)\\ \nonumber
     & =  \sum\limits_{k'\in \mathbb{K}} \eta_{k'}  \left(         
    \hat{\mathbf{h}}_{k0}^H \boldsymbol{\Theta}_{k'0} \hat{\mathbf{h}}_{k0}  + \sum\nolimits_{l\in \mathbb{L}} \hat{\mathbf{h}}_{kl}^H \boldsymbol{\Theta}_{k'l}  \hat{\mathbf{h}}_{kl} 
      \right),
\end{align}
where the derivations $(a)$ follows from the orthonormality of symbols, i.e., $\mathbb{E}[x_{k'}^*x_{k}]=0$ for $k'\neq k$, and $\mathbb{E}[|x_{k}|^2]=1$ for any $k$; $(b)$ uses the fact that the variance of a sum of independent random variables equals the sum of their variances; and $(c)$ is based on the independence between channel estimates and their estimation errors.

Similarly, the variance of $\mathcal{I}_2$ is computed as
\begin{align}  \label{eQN_deriv0001} \nonumber
     \mathbb{E}\left[|\mathcal{I}_2|^2\right] & =  \sum\limits_{k'\in \mathbb{K}\backslash \{k\}} \eta_{k'} \mathbb{E}\left[\left| \hat{\mathbf{h}}_{k0}^H \hat{\mathbf{h}}_{k'0} + \sum\nolimits_{l\in \mathbb{L}}\hat{\mathbf{h}}_{kl}^H \hat{\mathbf{h}}_{k'l} \right|^2 \right] \\ 
     & =  \sum\limits_{k'\in \mathbb{K}\backslash \{k\}} \eta_{k'} \left| \hat{\mathbf{h}}_{k0}^H \hat{\mathbf{h}}_{k'0} + \sum\nolimits_{l\in \mathbb{L}}\hat{\mathbf{h}}_{kl}^H \hat{\mathbf{h}}_{k'l} \right|^2, 
\end{align}
due to the deterministic nature of the channel estimates.
The variance of $\mathcal{I}_3$ equals to 
\begin{align}      
    \mathbb{E}\left[|\mathcal{I}_3|^2\right] & = \frac{1}{p_u}\mathbb{E}\left[\left |\hat{\mathbf{h}}_{k0}^H\mathbf{z}_0 + \sum\nolimits_{l\in \mathbb{L}}\hat{\mathbf{h}}_{kl}^H\mathbf{z}_l \right |^2\right] \\ \nonumber
    & = \frac{1}{p_u}\mathbb{E}\left[\left |\hat{\mathbf{h}}_{k0}^H\mathbf{z}_0\right |^2\right] + \sum\nolimits_{l\in \mathbb{L}}\mathbb{E}\left[\left |\hat{\mathbf{h}}_{kl}^H\mathbf{z}_l \right |^2\right] \\ \nonumber
    & = \frac{1}{p_u} \left (\hat{\mathbf{h}}_{k0}^H \mathbb{E}\left[\mathbf{z}_0\mathbf{z}_0^H\right]\hat{\mathbf{h}}_{k0} + \sum\limits_{l\in \mathbb{L}} \hat{\mathbf{h}}_{kl}^H \mathbb{E}\left[\mathbf{z}_l \mathbf{z}_l^H\right]  \hat{\mathbf{h}}_{kl} \right ) \\ \nonumber
    & = \frac{\sigma_z^2}{p_u} \left ( \hat{\mathbf{h}}_{k0}^H  \mathbf{I}_{N_b} \hat{\mathbf{h}}_{k0} + \sum\nolimits_{l\in \mathbb{L}} \hat{\mathbf{h}}_{kl}^H \mathbf{I}_{N_a}  \hat{\mathbf{h}}_{kl} \right)\\ \nonumber
    & = \frac{\sigma_z^2}{p_u} \left ( \|  \hat{\mathbf{h}}_{k0}\|^2 + \sum\nolimits_{l\in \mathbb{L}} \| \hat{\mathbf{h}}_{kl}\|^2 \right).
\end{align}
Substituting these variances into \eqref{cfmmimo:formularSNR} yields \eqref{eQN_sinrULcentralized}. 
\end{IEEEproof}

\begin{algorithm}
\SetAlgoLined \label{Algorithm_UL}
\DontPrintSemicolon
\KwIn{$\left\{ \hat{\mathbf{h}}_{k0}, \hat{\mathbf{h}}_{kl}, \boldsymbol{\Theta}_{k0}, \boldsymbol{\Theta}_{kl}\right\}_{k\in\mathbb{K}, l\in\mathbb{L}}$,  $p_u$, $\sigma_z^2$, $\epsilon$;  }
\KwOut{Optimal power coefficients $\{\eta_k^*\}_{k\in\mathbb{K}}$ }
Initialization: \;
$t \leftarrow 0$\;
$\gamma^{(0)}_\text{low}\leftarrow 0$,\quad $\gamma^{(0)}_\text{high}\leftarrow \max\left\{ \frac{\left( \|\hat{\mathbf{h}}_{k0}\|^2+ \sum_{l\in \mathbb{L}} \| \hat{\mathbf{h}}_{kl} \|^2 \right)p_u}{\sigma_z^2}\right\}$\;
\While{$\gamma^{(t)}_\text{high} - \gamma^{(t)}_\text{low} > \epsilon$}{
    $\gamma_t \leftarrow \frac{1}{2}(\gamma^{(t)}_\text{low} + \gamma^{(t)}_\text{high})$\;   
    \textbf{Convex Feasibility Check}:  
    \begin{equation*}
    \begin{aligned}
        \text{Find}\quad & \{\eta_k\}_{k\in\mathbb{K}} \\
        \text{s.t.}\quad & \gamma_k^{hcf} \geq \gamma_t,\: \forall k \in \mathbb{K} \\
        & 0\leq \eta_{k} \leq 1, \quad \forall k \in \mathbb{K}
    \end{aligned}
    \end{equation*}    
    \If{feasible}{
        $\gamma_{\text{low}}^{(t+1)} \leftarrow \gamma_t$,\quad 
        $\gamma_{\text{high}}^{(t+1)} \leftarrow \gamma_{\text{high}}^{(t)}$\;
        $\eta_k^* \leftarrow \eta_k$\;
    }
    \Else{
        $\gamma_{\text{low}}^{(t+1)} \leftarrow \gamma_{\text{low}}^{(t)}$, \quad
        $\gamma_{\text{high}}^{(t+1)} \leftarrow \gamma_t$\;
    }
    $t \leftarrow t + 1$\;
}
\Return  $\{\eta_k^*\}_{k\in\mathbb{K}}$ \;
\caption{Uplink HCF Max-Min Power Control}
\end{algorithm}
\subsection{Max-Min Uplink Power Optimization}
Max-min power control has proven highly effective in enhancing fairness for CF systems \cite{Ref_ngo2017cellfree, Ref_bashar2019uplink}. Building on this insight, we design a HCF max-min fairness algorithm to dynamically adjust UE's transmit power in the uplink. The optimization formula can be represented as
\begin{equation}  
\begin{aligned} \label{eqnIRS:optimizationMRTvector}
\max_{\{\eta_k\}_{k\in\mathbb{K}}} \min_{k}\quad &  \gamma_{k}^{hcf}  \\
\textrm{s.t.} \quad & 0\leqslant \eta_{k} \leqslant 1,\quad \forall k\in \mathbb{K}
\end{aligned}.
\end{equation}
Introducing a slack variable $\gamma_t$, which represents the minimum SINR across users, the problem is reformulated as 
\begin{equation}  
\begin{aligned} \label{EQN_maxminOptim}
\max_{\{\eta_k\}_{k\in\mathbb{K}},\;\gamma_t } \quad &  \gamma_t \\
\textrm{s.t.} \quad & \gamma_{k}^{hcf} \geqslant \gamma_t, \quad \forall k\\
\quad & 0\leqslant \eta_{k} \leqslant 1,\quad \forall k\in \mathbb{K}.
\end{aligned}
\end{equation}

Since \( \gamma_{k}^{hcf} \) is quasiconcave with respect to \( \{\eta_k\}_{k \in \mathbb{K}} \), the constraint set \( \gamma_{k}^{hcf} \geq \gamma_t \) is convex. Therefore, for a fixed \( \gamma_t \), the constraint domain in \eqref{EQN_maxminOptim} is convex, making it straightforward to determine whether a given \( \gamma_t \) is feasible. Consequently, the bisection method is applied: first, we select an interval \( (\gamma_{low}, \gamma_{high}) \) that contains the optimal value \( \gamma_t^* \). Then, we check the feasibility of the midpoint \( \gamma_t = \frac{\gamma_{low} + \gamma_{high}}{2} \). If \( \gamma_t \) is feasible, the search interval is updated to \( (\gamma_t, \gamma_{high}) \); otherwise, it is updated to \( (\gamma_{low}, \gamma_t) \). This process repeats until the search interval is sufficiently small, less than a tolerance $\epsilon>0$, as depicted in Algorithm~\ref{Algorithm_UL}.

\subsection{Cellular Massive MIMO}
To facilitate comparison, we analyze cellular systems, deriving its closed-form SE expression under spatially correlated channels and pilot contamination. In the uplink of a cellular system, the BS observes 
\begin{equation} \label{eqnUPLINKmodel} 
    \mathbf{y}_b = \sqrt{p_u} \sum\nolimits_{k\in \mathbb{K}} \sqrt{\eta_k} \mathbf{g}_k x_k + \mathbf{z}_b,
\end{equation}
where $\mathbf{g}_k \in \mathbb{C}^{M}$ represents the channel vector between the BS antennas and user $k$, and the receiver noise $\mathbf{z}_b \sim \mathcal{CN}(\mathbf{0},\sigma_z^2\mathbf{I}_{M})$. 
Let $\hat{\mathbf{g}}_k$ be the MMSE estimate of $\mathbf{g}_k$, and the estimation error $\tilde{\mathbf{g}}_k = \mathbf{g}_k - \hat{\mathbf{g}}_k$ follows a complex Gaussian distribution, $\tilde{\mathbf{g}}_k \sim \mathcal{CN}(\mathbf{0}, \boldsymbol{\Theta}_k)$. The covariance matrix $\boldsymbol{\Theta}_k$ is 
\begin{equation}
    \boldsymbol{\Theta}_k = \mathbb{E} \left[ \tilde{\mathbf{g}}_k \tilde{\mathbf{g}}_k^H \right] = \mathbf{R}_k - p_u \tau_p \mathbf{R}_k \boldsymbol{\Gamma}_k^{-1} \mathbf{R}_k,
\end{equation}
where $\mathbf{R}_k = \mathbb{E} \left[ \mathbf{g}_k \mathbf{g}_k^H \right]$ is the spatial correlation matrix, and $
    \boldsymbol{\Gamma}_k = p_u \tau_p \sum_{k' \in \mathcal{P}_k} \mathbf{R}_{k'} + \sigma_z^2 \mathbf{I}_M
$.
The BS correlates $\mathbf{y}_b$ with $\hat{\mathbf{g}}_k$,  to generate a soft estimate of $x_k$, as follows:
\begin{align} \label{massiveMIMO:MFsoftestimateUL} 
    &\hat{x}_k = \frac{1}{\sqrt{p_u}}\hat{\mathbf{g}}_k^H \mathbf{y}_b 
    %&= \sqrt{ \eta_k} \hat{\mathbf{g}}_k^H \hat{\mathbf{g}}_k  x_k  + \sqrt{ \eta_k} \hat{\mathbf{g}}_k^H \tilde{\mathbf{g}}_k  x_k \\&+ \sum\limits_{k'\in \mathbb{K}\backslash \{k\}} \sqrt{\eta_{k'}} \hat{\mathbf{g}}_k^H \mathbf{g}_{k'}  x_{k'} +  \frac{\hat{\mathbf{g}}_k^H\mathbf{z}}{\sqrt{p_u}}.
    \\ \nonumber
    &= \sqrt{ \eta_k} \hat{\mathbf{g}}_k^H (\hat{\mathbf{g}}_k + \tilde{\mathbf{g}}_k) x_k   + \sum\limits_{k'\in \mathbb{K}\backslash \{k\}} \sqrt{\eta_{k'}} \hat{\mathbf{g}}_k^H \mathbf{g}_{k'}  x_{k'} +  \frac{\hat{\mathbf{g}}_k^H\mathbf{z}_b}{\sqrt{p_u}}.
\end{align}

\begin{proposition}
The achievable SE for user $k$ in the uplink of a cellular system, considering correlated channels and pilot contamination, is expressed as $
    R_k^{c}=(1-\frac{\tau_p}{\tau_u}) \mathbb{E} [\log_2(1+\gamma_k^{c}) ]$,
where the instantaneous effective SINR is
\begin{equation} \label{eQN_sinr_UL_cellular}
    \gamma_k^{c} = \frac{ \eta_k \|  \hat{\mathbf{g}}_k \|^4  }{  \hat{\mathbf{g}}_k^H \left( \sum\limits_{k'\in \mathbb{K}\backslash \{k\}} \eta_{k'}  \hat{\mathbf{g}}_{k'}\hat{\mathbf{g}}_{k'}^H  + \sum\limits_{k'\in \mathbb{K}} \eta_{k'} \boldsymbol{\Theta }_{k'} +\frac{\sigma_z^2}{p_u}\mathbf{I}_M\right) \hat{\mathbf{g}}_k  }.
\end{equation}
\end{proposition}
\subsection{Cell-Free Massive MIMO}
We derive closed-form SE expression for CF systems under spatially correlated channels and pilot contamination. This goes beyond previous works, which were primarily limited to single-antenna APs or independent channels.
In a CF system with $A=\frac{M}{N_a}$ distributed APs, each equipped with $N_a$ antennas, the received signal at AP $a$, where $a \in \mathbb{A}=\{1,\ldots, A\} $, is expressed by
\begin{equation}  
    \mathbf{y}_a = \sqrt{p_u} \sum\nolimits_{k\in \mathbb{K}} \sqrt{\eta_k} \mathbf{g}_{ka} x_k + \mathbf{z}_a.
\end{equation}
Here, the receiver noise $\mathbf{z}_a \sim \mathcal{CN}(\mathbf{0},\sigma_z^2\mathbf{I}_{N_a})$, and $\mathbf{g}_{ka} \in \mathbb{C}^{N_a}$ denotes the channel vector between AP $a$ and user $k$. Similarly, the MMSE estimate of $\mathbf{g}_{ka}$ is denoted as $\hat{\mathbf{g}}_{ka}$, while the estimation error $\tilde{\mathbf{g}}_{ka}=\mathbf{g}_{ka} - \hat{\mathbf{g}}_{ka}$ follows $\mathcal{CN}(\mathbf{0}, \boldsymbol{\Theta }_{ka} )$. Its covariance matrix $
    \boldsymbol{\Theta}_{ka} = \mathbf{R}_{ka} - p_u \tau_p \mathbf{R}_{ka} \boldsymbol{\Gamma}_{ka}^{-1} \mathbf{R}_{ka}$, with $\mathbf{R}_{ka}=\mathbb{E}[\mathbf{g}_{ka}\mathbf{g}_{ka}^H ] $ and $
    \boldsymbol{\Gamma}_{ka} = p_u \tau_p \sum_{k' \in \mathcal{P}_k} \mathbf{R}_{k'a} + \sigma_z^2 \mathbf{I}_{N_a}
$.    
For a fair comparison, we consider that the CPU performs centralized MF combining based on the received signals forwarded by the APs, resulting in a soft estimate of
\begin{align}  \label{eQx_UL_rxSigCellular}
    \hat{x}_k =& \frac{1}{\sqrt{p_u}} \sum\nolimits_{a\in \mathbb{A}}\hat{\mathbf{g}}_{ka}^H\mathbf{y}_a\\ \nonumber =& \sum\nolimits_{a\in \mathbb{A}}\hat{\mathbf{g}}_{ka}^H\left(  \sum\nolimits_{k'\in \mathbb{K}} \sqrt{\eta_{k'}} \mathbf{g}_{k'a} x_{k'} + \frac{1}{\sqrt{p_u}}\mathbf{z}_a \right) \\ \nonumber
    =& \sqrt{ \eta_k}  \sum_{a\in \mathbb{A}}\hat{\mathbf{g}}_{ka}^H \hat{\mathbf{g}}_{ka} x_k  +   \sum\nolimits_{k'\in \mathbb{K}} \sqrt{\eta_{k'}}  \sum_{a\in \mathbb{A}}\hat{\mathbf{g}}_{ka}^H \tilde{\mathbf{g}}_{k'a}  x_{k'} \\ +& \sum\nolimits_{k'\in \mathbb{K}\backslash \{k\}} \sqrt{\eta_{k'}}  \sum_{a\in \mathbb{A}}\hat{\mathbf{g}}_{ka}^H \hat{\mathbf{g}}_{k'a}   x_{k'}+  \frac{1}{\sqrt{p_u}}\sum_{a\in \mathbb{A}}\hat{\mathbf{g}}_{ka}^H\mathbf{z}_a.  \nonumber
\end{align}
\begin{proposition}
The achievable SE for user $k$ in the uplink of a cell-free system, considering correlated channels and pilot contamination, is given by $
    R_k^{cf}=(1-\frac{\tau_p}{\tau_u}) \mathbb{E} [\log_2(1+\gamma_k^{cf}) ]$ with
 the effective SINR of
\begin{equation} \label{eQN_sinr_UL_cellular}
    \gamma_k^{cf} = \frac{ \eta_k \left(\sum\nolimits_{a\in \mathbb{A}} \|  \hat{\mathbf{g}}_{ka} \|^2 \right)^2  }{ \left\{ \begin{aligned}    &  \sum\nolimits_{k'\in \mathbb{K}}  \eta_{k'} \sum \nolimits_{a\in \mathbb{A}} \hat{\mathbf{g}}_{ka}^H\boldsymbol{\Theta}_{k'a}\hat{\mathbf{g}}_{ka} \\
    &+ 
    \sum\limits_{k'\in \mathbb{K}\backslash \{k\}} \eta_{k'} \left|  \sum\limits_{a\in \mathbb{A}}\hat{\mathbf{g}}_{ka}^H \hat{\mathbf{g}}_{k'a} \right|^2  + \frac{\sigma_z^2}{p_u}  \sum \limits_{a\in \mathbb{A}} \| \hat{\mathbf{g}}_{ka}\|^2
       \end{aligned} \right\}
    }.
\end{equation}
\end{proposition}

\section{Downlink Data Transmission}
In the downlink, the information-bearing symbols intended for $K$ users are modeled as zero-mean, unit-variance, and independent random variables. These symbols are collectively represented as $\mathbf{u}=[u_1,\ldots,u_K]^T$, where  $\mathbb{E}[\mathbf{u}\mathbf{u}^H]=\mathbf{I}_K$.  

\subsection{Linear Precoding for HCF Massive MIMO}

\begin{figure*}
\begin{align}   \label{eQN_SINR_DL_MR_final}
    \xi_k^{hcf}  & = \frac{  \left( \sqrt{p_b\eta_{k0} p_u \tau_p\mathrm{tr}\left(\mathbf{R}_{k0} \boldsymbol{\Gamma}_{k0}^{-1}\mathbf{R}_{k0}\right)} + \sum\nolimits \nolimits_{l\in\mathbb{L}} \sqrt{p_a\eta_{kl}p_u \tau_p\mathrm{tr}\left(\mathbf{R}_{kl} \boldsymbol{\Gamma}_{kl}^{-1}\mathbf{R}_{kl}\right)} \right)^2  }{ \left\{ \begin{aligned} &       
     p_b \sum\nolimits_{k'\in \mathcal{P}_k\backslash \{k\}} \eta_{k'0}  \frac{p_u \tau_p \left| \mathrm{tr}(     \mathbf{R}_{k0}  \boldsymbol{\Gamma}_{k'0}^{-1}\mathbf{R}_{k'0} )  \right|^2 
    } {\mathrm{tr}\left(\mathbf{R}_{k'0} \boldsymbol{\Gamma}_{k'0}^{-1}\mathbf{R}_{k'0}\right)   }   +  p_b \sum\nolimits_{{k'}\in \mathbb{K}}  \eta_{k'0}  \frac{  \tr \left ( \mathbf{R}_{k0} \mathbf{R}_{k'0} \boldsymbol{\Gamma}_{k'0}^{-1}\mathbf{R}_{k'0}    \right)} {  \tr\left(\mathbf{R}_{k'0} \boldsymbol{\Gamma}_{k'0}^{-1}\mathbf{R}_{k'0}\right) }   \\ &  + p_a  \sum\limits_{k'\in \mathcal{P}_k \backslash \{k\}}\sum\nolimits_{l\in\mathbb{L}} \eta_{k'l}  \frac{p_u \tau_p \left| \mathrm{tr}(     \mathbf{R}_{kl}  \boldsymbol{\Gamma}_{k'l}^{-1}\mathbf{R}_{k'l} )  \right|^2 
    } {\mathrm{tr}\left(\mathbf{R}_{k'l} \boldsymbol{\Gamma}_{k'l}^{-1}\mathbf{R}_{k'l}\right)   }    + p_a   \sum\nolimits_{{k'}\in \mathbb{K}}  \sum\nolimits_{l\in\mathbb{L}} \eta_{k'l}  \frac{  \tr \left ( \mathbf{R}_{kl} \mathbf{R}_{k'l} \boldsymbol{\Gamma}_{k'l}^{-1}\mathbf{R}_{k'l}    \right)} {  \tr\left(\mathbf{R}_{k'l} \boldsymbol{\Gamma}_{k'l}^{-1}\mathbf{R}_{k'l}\right) }      +\sigma_z^2 \end{aligned} \right\} 
    }
\end{align} 
\end{figure*}

Let $\mathbf{w}_{k0}\in \mathbb{C}^{N_b}$ denote the precoding vector for user $k$ at the cBS. TDD channel reciprocity allows the cBS to utilize uplink channel estimates for downlink precoding \cite{Ref_marzetta2015massive}. To spatially multiplex symbols, the cBS employs conjugate beamforming (CBF) \cite{Ref_ngo2017cellfree}, i.e., $\mathbf{w}_{k0}=\hat{\mathbf{h}}^*_{k0}/ \sqrt{ \mathbb{E}[\|\hat{\mathbf{h}}_{k0}\|^2]} $, where the scaling is used to meet the normalization condition $\mathbb{E}[\|\mathbf{w}_{k0}\|^2]= 1$.
Consequently, the cBS transmits 
\begin{equation} \label{GS_TxSignlCBS_S0}
    \mathbf{s}_0 = \sqrt{p_b}\sum_{k\in \mathbb{K}} \sqrt{\eta_{k0}} \mathbf{w}_{k0} u_k= \sqrt{p_b}\sum_{k\in \mathbb{K}}  \frac{\sqrt{\eta_{k0}}  \hat{\mathbf{h}}^*_{k0} u_k }{ \sqrt{ \mathbb{E}[\|\hat{\mathbf{h}}_{k0}\|^2]}  }  ,
\end{equation} where $\eta_{k0}$ indicates the power coefficient for user $k$ at the cBS, and $p_b$ denotes the maximum transmit power of the cBS.  To meet the average power constraint, i.e., $\mathbb{E}[\|\mathbf{s}_0\|^2]\leqslant p_b$, the summation of coefficients must satisfy $\sum_{k\in \mathbb{K}} \eta_{k0}\leqslant 1$. 

Similarly, let $\mathbf{w}_{kl}\in \mathbb{C}^{N_a}$ denote the precoding vector assigned to $u_k$ at AP $l$. 
The cBS uses $\mathbf{w}_{kl}=\hat{\mathbf{h}}^*_{kl}/\sqrt{ \mathbb{E}[\|\hat{\mathbf{h}}_{kl}\|^2]}$ to generate the transmit signal for eAP $l$, i.e.,
\begin{equation} \label{eQn:compositeTxSig_MR}
    \mathbf{s}_l = \sqrt{p_a}\sum_{k\in \mathbb{K}} \sqrt{\eta_{kl}}\mathbf{w}_{kl} u_k = \sqrt{p_a} \sum_{k\in \mathbb{K}} \frac{\sqrt{\eta_{kl}} \hat{\mathbf{h}}^*_{kl} u_k} {\sqrt{ \mathbb{E}[\|\hat{\mathbf{h}}_{kl}\|^2]} } ,
\end{equation}
where $p_a$ stands for the maximum transmit power of eAPs, and $\eta_{kl}$ represents the power coefficient for eAP $l$ to user $k$, subject to the constraint $\sum_{k\in \mathbb{K}} \eta_{kl}\leqslant 1$ in order to satisfy $\mathbb{E}[\|\mathbf{s}_l\|^2]\leqslant p_a$. 
Sending $\{\mathbf{s}_l\}_{l\in \mathbb{L}}$ to the respective eAPs results in the transmission of \(N_a\times L\) complex-valued scalars over the fronthaul network.
With noise $n_k\sim \mathcal{CN}(0,\sigma^2_z)$, the observation at user $k$ is given by  $y_k  =  \mathbf{h}_{k0}^T\mathbf{s}_0 +  \sum\nolimits_{l\in \mathbb{L} } \mathbf{h}_{kl}^T\mathbf{s}_l  +n_k  $, which can be further expanded as
\begin{align}   \label{eQn_downlinkModel} 
   y_k =  & \sqrt{p_b}\sum\nolimits_{k'\in \mathbb{K}} \sqrt{\eta_{k'0}} \mathbf{h}_{k0}^T \frac{  \hat{\mathbf{h}}^*_{k'0} }{ \sqrt{ \mathbb{E}[\|\hat{\mathbf{h}}_{k'0}\|^2]}  } u_{k'}\\  \nonumber
     + & \sqrt{p_a}  \sum\nolimits_{k'\in \mathbb{K}} \sum\nolimits_{l\in\mathbb{L}} \sqrt{\eta_{k'l}} \mathbf{h}_{kl}^T \frac{ \hat{\mathbf{h}}^*_{k'l} } {\sqrt{ \mathbb{E}[\|\hat{\mathbf{h}}_{k'l}\|^2]} } u_{k'} +n_k.
\end{align}

In massive MIMO, users typically do not know channel estimates due to the absence of downlink pilots \cite{ngo2017no}, rendering coherent detection unfeasible.  As a result, user $k$ must rely on channel statistics $ \mathbb{E} [ \| \hat{\mathbf{h}}_{k0} \|^2  ]$ and $ \{ \mathbb{E} [ \| \hat{\mathbf{h}}_{kl} \|^2  ] \}_{l\in \mathbb{L}}$ instead of channel estimates $\hat{\mathbf{h}}_{k0}$ and $\{\hat{\mathbf{h}}_{kl}\}_{l\in \mathbb{L}}$ to detect $y_k$. This introduces an additional performance loss, termed the \textit{channel uncertainty error}, as marked by $\mathcal{J}_1$ in \eqref{eQn_DLGeneralSig}. To facilitate the derivation of the closed-form SE expression, \eqref{eQn_downlinkModel} is decomposed into 
\begin{align} \nonumber \label{eQn_DLGeneralSig}
    y_k = &  \underbrace{ \left( \frac{\sqrt{p_b\eta_{k0}}\mathbb{E} [   \hat{\mathbf{h}}_{k0}^T \hat{\mathbf{h}}_{k0}^* ]} { \sqrt{ \mathbb{E}[\|\hat{\mathbf{h}}_{k0}\|^2]} }   + \sum\nolimits_{l\in\mathbb{L}}  \frac{ \sqrt{p_a \eta_{kl}} \mathbb{E} [ \hat{\mathbf{h}}_{kl}^T\hat{\mathbf{h}}_{kl}^*]} { \sqrt{ \mathbb{E}[\|\hat{\mathbf{h}}_{kl}\|^2]} } \right) u_k}_{\mathcal{S}_1:\:\text{desired signal over channel statistics}} \\ + & \underbrace{ \left( \begin{aligned}
        &  \frac{ \sqrt{p_b\eta_{k0}}\left( \mathbf{h}_{k0}^T \hat{\mathbf{h}}_{k0}^* - \mathbb{E} [   \hat{\mathbf{h}}_{k0}^T \hat{\mathbf{h}}_{k0}^* ] \right)} { \sqrt{ \mathbb{E}[\|\hat{\mathbf{h}}_{k0}\|^2]} }   \\ \nonumber & + \sqrt{p_a}\sum\nolimits_{l\in\mathbb{L}}  \frac{ \sqrt{\eta_{kl}}\left( \mathbf{h}_{kl}^T\hat{\mathbf{h}}_{kl}^* - \mathbb{E} [ \hat{\mathbf{h}}_{kl}^T\hat{\mathbf{h}}_{kl}^*] \right) } { \sqrt{ \mathbb{E}[\|\hat{\mathbf{h}}_{kl}\|^2]} }   \end{aligned} \right) u_k  }_{\mathcal{J}_1:\:channel\:uncertainty\:error} \\  + &
    \underbrace{  \left ( \begin{aligned} &\sqrt{p_b}\sum\nolimits_{k'\in \mathbb{K} \backslash \{k\}}  \frac{ \sqrt{\eta_{k'0}} \mathbf{h}_{k0}^T \hat{\mathbf{h}}^*_{k'0} }{ \sqrt{ \mathbb{E}[\|\hat{\mathbf{h}}_{k'0}\|^2]}  } \\  
     + & \sqrt{p_a}  \sum\nolimits_{k'\in \mathbb{K} \backslash \{k\} } \sum\nolimits_{l\in\mathbb{L}}  \frac{ \sqrt{\eta_{k'l}} \mathbf{h}_{kl}^T\hat{\mathbf{h}}^*_{k'l} } {\sqrt{ \mathbb{E}[\|\hat{\mathbf{h}}_{k'l}\|^2]} }   \end{aligned}\right) u_{k'}
    }_{\mathcal{J}_2:\:inter-user\:interference}+ n_k. 
\end{align}
\begin{proposition} \label{prop_dl_hcf_se}
The achievable SE for user $k$ in the downlink of HCF massive MIMO systems is expressed as $C_k^{hcf}=\mathbb{E} \Bigl[\log_2(1+\xi_k^{hcf}) \Bigr]$,
where the effective SINR is given in \eqref{eQN_SINR_DL_MR_final}.
\end{proposition}
\begin{IEEEproof}
The proof is detailed in Appendix \eqref{app_proposition_DLHCF}.
\end{IEEEproof}

\subsection{Joint Max-Min Downlink Power Optimization}

Given the downlink SINR expression in \eqref{eQN_SINR_DL_MR_final}, we aim to design a max-min fairness algorithm for joint power control of the cBS and APs. The problem is formulated as \cite{Ref_bashar2019uplink}:
\begin{align}   \label{eq:maxmin_problem}
    \max_{\{\eta_{k0}, \eta_{kl}\}_{k\in\mathbb{K},l\in\mathbb{L}}} & \min_{k \in \mathbb{K}} \quad  \xi_k^{hcf}\\
    \text{s.t.} \quad  & \sum\nolimits_{k\in\mathbb{K}} \eta_{k0} \leq 1, \; \sum\nolimits_{k\in\mathbb{K}} \eta_{kl} \leq 1, \ \forall l \in \mathbb{L}, \nonumber \\
    & \eta_{k0} \geq 0, \quad \eta_{kl} \geq 0, \quad \forall k \in \mathbb{K}, l \in \mathbb{L}. \nonumber
\end{align}
The numerator of $\xi_k^{hcf}$ consists of a sum of square root terms, making it a concave function. To facilitate optimization, we introduce auxiliary variables $\nu_{k0} = \sqrt{\eta_{k0}}$ and $\nu_{kl} = \sqrt{\eta_{kl}}$, collectively denoted as $\boldsymbol{\nu} = \{\nu_{k0}, \nu_{kl}\}_{k\in\mathbb{K},l\in\mathbb{L}}$. 

Introducing a slack variable $\xi_t$ to reformulate \eqref{eq:maxmin_problem} into 
\begin{align}
    \label{eq:transformed_problem}
    \max_{\boldsymbol{\nu},\, \xi_t} \quad & \xi_t  \\
    \text{s.t.} \quad & \left(\nu_{k0} A_{k0} + \sum\nolimits_{l\in\mathbb{L}} \nu_{kl} A_{kl} \right)^2 \geq \xi_t \mathcal{D}_k(\boldsymbol{\nu}), \tag{C1} \nonumber \\
    & \sum\nolimits_{k\in\mathbb{K}} \nu_{k0}^2 \leq 1, \quad \sum\nolimits_{k\in\mathbb{K}} \nu_{kl}^2 \leq 1, \quad \forall l \in \mathbb{L}, \nonumber\\
    & \nu_{k0} \geq 0, \quad \nu_{kl} \geq 0, \quad \forall k \in \mathbb{K}, l \in \mathbb{L}. \nonumber
\end{align}
We define: $A_{k0} = \sqrt{ p_bp_u \tau_p \tr(\mathbf{R}_{k0} \boldsymbol{\Gamma}_{k0}^{-1}\mathbf{R}_{k0})}$ and $A_{kl} = \sqrt{ p_a p_u \tau_p \tr(\mathbf{R}_{kl} \boldsymbol{\Gamma}_{kl}^{-1}\mathbf{R}_{kl})}$, for ease of notation. The denominator of $\xi_k^{hcf}$ is rewritten as
    \begin{align} \nonumber
        \mathcal{D}_k(\boldsymbol{\nu})&=\sum\nolimits_{k'\in \mathcal{P}_k\backslash \{k\}} \left(\nu^2_{k'0} B^2_{k0,k'} + \sum\nolimits_{l\in\mathbb{L}}\nu^2_{k'l} B^2_{kl,k'}\right) \\ &+ \sum\limits_{{k'}\in \mathbb{K}} \left( \nu^2_{k'0} C^2_{k0,k'} + \sum\nolimits_{l\in\mathbb{L}}\nu^2_{k'l} C^2_{kl,k'} \right) + \sigma_z^2,
    \end{align}
where 
\begin{align} \label{EQN_coefficientsinmaxmin} 
    B_{k0,k'} &=  \sqrt{\frac{p_bp_u \tau_p \left| \mathrm{tr}(     \mathbf{R}_{k0}  \boldsymbol{\Gamma}_{k'0}^{-1}\mathbf{R}_{k'0} )  \right|^2 
    } {\mathrm{tr}\left(\mathbf{R}_{k'0} \boldsymbol{\Gamma}_{k'0}^{-1}\mathbf{R}_{k'0}\right)   }}
    \\ \nonumber
    B_{kl,k'} &=  \sqrt{ \frac{p_ap_u \tau_p \left| \mathrm{tr}(     \mathbf{R}_{kl}  \boldsymbol{\Gamma}_{k'l}^{-1}\mathbf{R}_{k'l} )  \right|^2 
    } {\mathrm{tr}\left(\mathbf{R}_{k'l} \boldsymbol{\Gamma}_{k'l}^{-1}\mathbf{R}_{k'l}\right)   } }
\end{align} and
\begin{align}
    C_{k0,k'} &= \sqrt{\frac{ p_b \tr \left ( \mathbf{R}_{k0} \mathbf{R}_{k'0} \boldsymbol{\Gamma}_{k'0}^{-1}\mathbf{R}_{k'0}    \right)} {  \tr\left(\mathbf{R}_{k'0} \boldsymbol{\Gamma}_{k'0}^{-1}\mathbf{R}_{k'0}\right) }}
    \\ \nonumber
    C_{kl,k'} &= \sqrt{\frac{ p_a \tr \left ( \mathbf{R}_{kl} \mathbf{R}_{k'l} \boldsymbol{\Gamma}_{k'l}^{-1}\mathbf{R}_{k'l}    \right)} {  \tr\left(\mathbf{R}_{k'l} \boldsymbol{\Gamma}_{k'l}^{-1}\mathbf{R}_{k'l}\right) }}.
\end{align}

The left-hand side of constraint (C1) in \eqref{eq:transformed_problem} is a convex quadratic function, while the right-hand side is convex in $\boldsymbol{\nu}^2$. Since the difference of two convex functions is generally non-convex, the optimization problem is non-convex. 
Given that $\mathcal{D}_k(\boldsymbol{\nu})$ is a sum of squared terms, we define a vector  \cite{Ref_ngo2017cellfree}
\begin{equation}
  \mathbf{f}_k(\boldsymbol{\nu}) = \left[ 
    \begin{aligned}
        & \left\{ \nu_{k'0} B_{k0,k'},\ \nu_{k'l} B_{kl,k'} \right\}_{\substack{k' \in \mathcal{P}_k \setminus \{k\},\: l \in \mathbb{L}}}\ \\
        & \left\{ \nu_{k'0} C_{k0,k'},\ \nu_{k'l} C_{kl,k'} \right\}_{\substack{k' \in \mathbb{K},\: l \in \mathbb{L}}} \\
        & \sigma_z 
    \end{aligned}
\right], 
\end{equation}
such that $\|\mathbf{f}_k(\boldsymbol{\nu})\|^2=\mathcal{D}_k(\boldsymbol{\nu})$. 
Finally, the constraint (C1) is reformulated as
\begin{equation}
    \nu_{k0} A_{k0} + \sum\nolimits_{l\in\mathbb{L}} \nu_{kl} A_{kl}  \geq \sqrt{\xi_t} \left\| \mathbf{f}_k(\boldsymbol{\nu}) \right\|,
\end{equation}
which is a standard second-order cone constraint. This transformation converts the problem into a convex form, allowing it to be efficiently solved through a sequence of convex feasibility problems.
To solve it, we present a bisection method, as detailed in Algorithm \ref{Algorithm_maxminDL}.

\begin{algorithm}
\SetAlgoLined \label{Algorithm_maxminDL}
\DontPrintSemicolon
\KwIn{$\left\{\mathbf{R}_{k0}, \mathbf{R}_{kl}, \boldsymbol{\Gamma}_{k0}, \boldsymbol{\Gamma}_{kl}\right\}_{k\in\mathbb{K}, l\in\mathbb{L}}$, $p_u$, $p_a$, $p_b$, $\sigma_z^2$, $\epsilon$; }
\KwOut{Optimal power coefficients $\boldsymbol{\nu}^*$}
Initialization: \;
$t \leftarrow 0$,\quad $\xi^{(0)}_\text{low}\leftarrow 0$ \; 
$\xi^{(0)}_\text{high} \leftarrow \max \left\{\frac{ \left( A_{k0} + \sum\nolimits_{l\in\mathbb{L}}  A_{kl} \right)^2   }{\sigma_z^2} \right\}$\; 
$\left\{A_{k0}, A_{kl}, B_{k0,k'}, B_{kl,k'}, C_{k0,k'}, C_{kl,k'}\right\}_{\forall k,k'\in\mathbb{K},l\in\mathbb{L}}$\;
\While{$\xi^{(t)}_\text{high} - \xi^{(t)}_\text{low} > \epsilon$}{
    $\xi_t \leftarrow \frac{1}{2}(\xi^{(t)}_\text{low} + \xi^{(t)}_\text{high})$\;   
    \textbf{Convex Feasibility Check}:  
    \begin{equation*}
    \begin{aligned}
        \text{Find}\quad & \boldsymbol{\nu} \\
        \text{s.t.}\quad & \nu_{k0} A_{k0} + \sum\limits_{l\in\mathbb{L}} \nu_{kl} A_{kl} \geq \sqrt{\xi_t} \|\mathbf{f}_k(\boldsymbol{\nu})\|,\: \forall k \in \mathbb{K} \\
        & \sum\nolimits_{k\in\mathbb{K}} \nu_{k0}^2 \leq 1,\quad\sum\nolimits_{k\in\mathbb{K}} \nu_{kl}^2 \leq 1, \quad \forall l \in \mathbb{L} \\
        & \nu_{k0} \geq 0,\; \nu_{kl} \geq 0, \quad \forall k \in \mathbb{K}, l \in \mathbb{L}
    \end{aligned} 
    \end{equation*}    
    \If{feasible}{
        $\xi_{\text{low}}^{(t+1)} \leftarrow \xi_t$,\quad 
        $\xi_{\text{high}}^{(t+1)} \leftarrow \xi_{\text{high}}^{(t)}$\;
        $\boldsymbol{\nu}^* \leftarrow \boldsymbol{\nu}$\;
    }
    \Else{
        $\xi_{\text{low}}^{(t+1)} \leftarrow \xi_{\text{low}}^{(t)}$, \quad
        $\xi_{\text{high}}^{(t+1)} \leftarrow \xi_t$\;
    }
    $t \leftarrow t + 1$\;
}
\Return $\boldsymbol{\nu}^*$\;
\caption{Downlink Max-Min Power Control}
\end{algorithm}

\subsection{Cellular Massive MIMO}
In the downlink of a cellular system, the BS transmits 
\begin{equation} \label{GS_TxSignlBS_S}
    \mathbf{s}_b = \sqrt{p_{c}}\sum\nolimits_{k\in \mathbb{K}}  \frac{\sqrt{\zeta_{k}}  \hat{\mathbf{g}}^*_{k} u_k }{ \sqrt{ \mathbb{E}[\|\hat{\mathbf{g}}_{k}\|^2]}  },
\end{equation}
where $p_c$ is the BS power constraint and $\zeta_{k}$ denotes power coefficient assigned to user $k$.
As a result, user $k$ sees
\begin{align}   
   y_k &= \mathbf{g}_{k}^T\mathbf{s}_b +n_k\\ \nonumber
   &= \sqrt{p_c}\sum\nolimits_{k'\in \mathbb{K}} \sqrt{\zeta_{k'}} \mathbf{g}_{k}^T \frac{  \hat{\mathbf{g}}^*_{k'} }{ \sqrt{ \mathbb{E}[\|\hat{\mathbf{g}}_{k'}\|^2]}  } u_{k'} +n_k.
\end{align}  
\begin{proposition}
The downlink SE for user $k$ in a cellular system, considering correlated channels and pilot contamination, is expressed as $C_k^{c}=\mathbb{E} \Bigl[\log_2(1+\xi_k^{c}) \Bigr]$,
where 
   \begin{align}   
    \xi_k^{c}  & = \frac{ p_u \tau_p  \zeta_{k} \mathrm{tr}\left(\mathbf{R}_{k} \boldsymbol{\Gamma}_{k}^{-1}\mathbf{R}_{k}\right)   }{ \left\{ \begin{aligned} &       
      \sum\nolimits_{k'\in \mathcal{P}_k\backslash \{k\}} \zeta_{k'}  \frac{p_u \tau_p \left| \mathrm{tr}(     \mathbf{R}_{k}  \boldsymbol{\Gamma}_{k'}^{-1}\mathbf{R}_{k'} )  \right|^2 
    } {\mathrm{tr}\left(\mathbf{R}_{k'} \boldsymbol{\Gamma}_{k'}^{-1}\mathbf{R}_{k'}\right)   } \\ &  +   \sum\nolimits_{{k'}\in \mathbb{K}}  \zeta_{k'}  \frac{  \tr \left ( \mathbf{R}_{k} \mathbf{R}_{k'} \boldsymbol{\Gamma}_{k'}^{-1}\mathbf{R}_{k'}    \right)} {  \tr\left(\mathbf{R}_{k'} \boldsymbol{\Gamma}_{k'}^{-1}\mathbf{R}_{k'}\right) }    + \frac{\sigma_z^2}{p_c} \end{aligned} \right\} 
    }.
\end{align} 
\end{proposition}

\subsection{Cell-Free Massive MIMO}
In the downlink of a CF system, AP $a$ transmits 
\begin{equation} 
    \mathbf{s}_a = \sqrt{p_a} \sum\nolimits_{k\in \mathbb{K}} \frac{\sqrt{\zeta_{ka}} \hat{\mathbf{g}}^*_{ka} u_k} {\sqrt{ \mathbb{E}[\|\hat{\mathbf{g}}_{ka}\|^2]} } ,
\end{equation}
where $\zeta_{ka}$ represents the power coefficient for AP $a$ to user $k$. The observation at user $k$ is given by  $y_k  =   \sum\nolimits_{a\in \mathbb{A} } \mathbf{g}_{ka}^T\mathbf{s}_a  +n_k  $, which can be further expanded as
\begin{align}   \label{eQn_downlinkModel} 
   y_k =  \sqrt{p_a}  \sum\limits_{k'\in \mathbb{K}} \sum\nolimits_{a\in\mathbb{A}} \sqrt{\zeta_{k'a}} \mathbf{g}_{ka}^T \frac{ \hat{\mathbf{g}}^*_{k'a} } {\sqrt{ \mathbb{E}[\|\hat{\mathbf{g}}_{k'a}\|^2]} } u_{k'} +n_k.
\end{align} 
\begin{proposition}
The downlink SE for user $k$ in a cell-free system, considering correlated channels and pilot contamination,  is expressed as $C_k^{cf}=\mathbb{E} \Bigl[\log_2(1+\xi_k^{cf}) \Bigr]$,
where 
   \begin{align}   
    \xi_k^{cf} & = \frac{  p_u \tau_p\left| \sum\nolimits \nolimits_{a\in\mathbb{A}} \sqrt{\zeta_{ka}\mathrm{tr}\left(\mathbf{R}_{ka} \boldsymbol{\Gamma}_{ka}^{-1}\mathbf{R}_{ka}\right)} \right|^2  }{ \left\{ \begin{aligned} &       
       \sum\limits_{k'\in \mathcal{P}_k \backslash \{k\}}\sum\limits_{a\in\mathbb{A}}   \frac{ \zeta_{k'a}p_u\tau_p \left| \mathrm{tr}(     \mathbf{R}_{ka}  \boldsymbol{\Gamma}_{k'a}^{-1}\mathbf{R}_{k'a} )  \right|^2 
    } {\mathrm{tr}\left(\mathbf{R}_{k'a} \boldsymbol{\Gamma}_{k'a}^{-1}\mathbf{R}_{k'a}\right)   }  \\ &  +   \sum\limits_{{k'}\in \mathbb{K}}  \sum\limits_{a\in\mathbb{A}}   \frac{ \zeta_{k'a} \tr \left ( \mathbf{R}_{ka} \mathbf{R}_{k'a} \boldsymbol{\Gamma}_{k'a}^{-1}\mathbf{R}_{k'a}    \right)} {  \tr\left(\mathbf{R}_{k'a} \boldsymbol{\Gamma}_{k'a}^{-1}\mathbf{R}_{k'a}\right) }      + \frac{\sigma_z^2}{p_a} \end{aligned} \right\} 
    } .
\end{align} 
\end{proposition}

\begin{comment}
\begin{table*}[t]
\centering
\caption{Summary of Major Simulation Parameters}
\label{tab:parameters}
\begin{tabularx}{\textwidth}{lXX}
\toprule
\textbf{Parameter} & \textbf{Microcell} & \textbf{Macrocell} \\
\midrule
Coverage radius & $500\,\mathrm{m}$ & $2000\,\mathrm{m}$ \\
Total antennas ($M$) & 128 & 384 \\
Active users ($K$) & 8 & 16 \\
Default HCF configuration & $N_b = 32$ antennas at cBS, $24$ eAPs ($4$ antennas each) & $N_b = 96$, $72$  eAPs ($4$ antennas each)\\
HCF-1/2 configuration & $N_b = 64$ antennas at cBS, $16$ eAPs ($4$ antennas each)  & $N_b = 192$, $48$ eAPs  ($4$ antennas each) \\
CF configuration & $32$ APs ($4$ antennas each) & $96$ APs ($4$ antennas each)\\
Cellular configuration & $128$-antenna BS  & $384$-antenna BS \\
Path loss model & 3GPP Urban Microcell & COST-Hata \\
Shadow fading ($\sigma_{\text{sd}}$) & $4\,\mathrm{dB}$ & $8\,\mathrm{dB}$ \\
Spatial correlation (angular deviation) & $30^\circ$ & $10^\circ$ \\
Pilot sequence length ($\tau_p$) & 4 & 8 \\
\midrule
UE power constraint ($p_u$) & \multicolumn{2}{c}{$200\,\mathrm{mW}$} \\
Per-antenna power constraint & \multicolumn{2}{c}{$50\,\mathrm{mW}$} \\
Noise power density & \multicolumn{2}{c}{$-174\,\mathrm{dBm/Hz}$} \\
Noise figure & \multicolumn{2}{c}{$9\,\mathrm{dB}$} \\
Bandwidth & \multicolumn{2}{c}{$5\,\mathrm{MHz}$} \\
Coherence block length ($\tau_c$) & \multicolumn{2}{c}{200} \\
\bottomrule
\end{tabularx}
\end{table*}
\end{comment}

\section{Performance Evaluation}
A numerical evaluation of the proposed HCF architecture is conducted, benchmarking its performance against CF and cellular systems with respect to both worst-case per-user SE and overall system capacity. Additionally, the results regarding the reduction in fronthaul costs and the saving of transmission power are presented.

\subsection{Simulation Configuration}
In our simulations, we examine two representative scenarios: a compact area reflecting a microcell-scale scenario and a larger area representing a macrocell-scale environment. 

\textbf{Microcell Scenario:} A circular area with a $500$-meter radius, consistent with the 3GPP Urban Microcell model, applies a path loss model:
\begin{equation}
\beta = -30.5 - 36.7 \lg\left( d \right)  + \mathcal{X},
\end{equation}
where $d$ is the propogation distance in meters, and shadowing $\mathcal{X}$ follows $ \mathcal{N}(0,\sigma_{sd}^2)$ with standard derivation $\sigma_{sd}=4\mathrm{dB}$.
The system deploys a total of $M=128$ antennas to serve $K=8$ active users. To implement HCF, by default, one-quarter of the antennas ($N_b=32$) are allocated to the cBS, along with $24$ eAPs, each equipped with four antennas, are uniformly distributed throughout the coverage area. This configuration reduces the fronthaul network scale by $25\%$, resulting in equivalent reductions in hardware costs and signaling overhead. 
We employ an HCF variant by increasing antenna centralization to half. In HCF-1/2, $64$ co-located antennas are placed at the cBS, with $16$ four-antenna eAPs. This setup further reduces the fronthaul scale to $50\%$ of the original CF network. For a fair comparison, all network configurations maintain the same number of antennas. Thus, the CF system consists of $32$ distributed APs, each equipped with four antennas, whereas in the cellular case, a BS with $128$ co-located antennas is used. At each simulation epoch, the locations of APs/eAPs and users are randomly varied.

\textbf{Macrocell Scenario:} The coverage radius extends to $2\mathrm{km}$ with $M=384$ antennas serving $K=16$ active users. The antenna distribution mirrors that of the microcell scenario, with each AP/eAP equipped with four antennas. 
The large-scale fading is given by $\beta=10^\frac{\mathcal{L}+\mathcal{X}}{10}$, where the shadowing $\mathcal{X}\sim \mathcal{N}(0,8^2)$, and path loss uses the COST-Hata model   \cite{Ref_ngo2017cellfree}:
\begin{equation} 
    \mathcal{L}= \begin{cases}
-L_0-35\lg(d), &  d>d_1 \\
-L_0-10\lg(d_1^{1.5}d^2), &  d_0<d\leq d_1 \\
-L_0-10\lg(d_1^{1.5}d_0^2), &  d\leq d_0
\end{cases},
\end{equation}
where the three-slope breakpoints  take values $d_0=10\mathrm{m}$ and $d_1=50\mathrm{m}$ while $L_0=140.72\mathrm{dB}$ in terms of 
\begin{IEEEeqnarray}{ll}
 L_0=46.3&+33.9\lg\left(f_c\right)-13.82\lg\left(h_{eAP}\right)\\ \nonumber
 &-\left[1.1\lg(f_c)-0.7\right]h_{UE}+1.56\lg\left(f_c\right)-0.8
\end{IEEEeqnarray}
with carrier frequency $f_c=1.9\mathrm{GHz}$, the height of eAP antenna $h_{eAP}=15\mathrm{m}$, and the height of UE $h_{UE}=1.65\mathrm{m}$. 

\textbf{General Parameters:} The UE transmit power is limited to $p_u=200\mathrm{mW}$. Each antenna has a power constraint of $50\,\mathrm{mW}$, meaning that a BS with $384$ antennas can transmit up to $19.2\,\mathrm{W}$, while each AP is restricted to a maximum power of $200\,\mathrm{mW}$. The system operates with a noise power spectral density of $-174\,\mathrm{dBm/Hz}$, a $9\,\mathrm{dB}$ noise figure, and a bandwidth of $5\,\mathrm{MHz}$. 
All antenna arrays are configured as half-wavelength-spaced uniform linear arrays. We simulate spatial correlation using the Gaussian local scattering model \cite[Sec.~2.6]{SIG-093}, applying an angular standard deviation of $10^\circ$ for macrocell scenarios and $30^\circ$ for microcell scenarios. Each coherence block contains $\tau_c = 200$ channel uses (e.g., achieved by $2\,\mathrm{ms}$ coherence time and $100\,\mathrm{kHz}$ coherence bandwidth). Without loss of generality, we set the length of the pilot sequence to half the total number of active users, i.e., $\tau_p = 4$ and $8$ in the microcell and macrocell scenarios, respectively. This implies that every two users share a pilot sequence, leading to pilot contamination.

\begin{figure*}[!tbph]
\centerline{
\subfloat[]{
\includegraphics[width=0.4\textwidth]{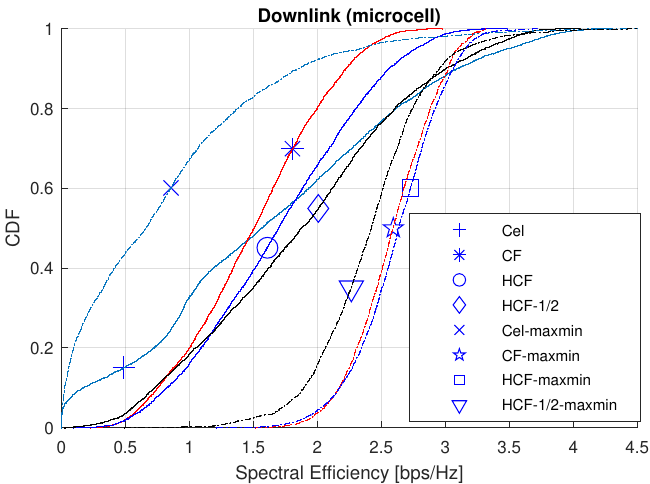}
\label{fig:result1} 
}
\hspace{20mm}
\subfloat[]{
\includegraphics[width=0.4\textwidth]{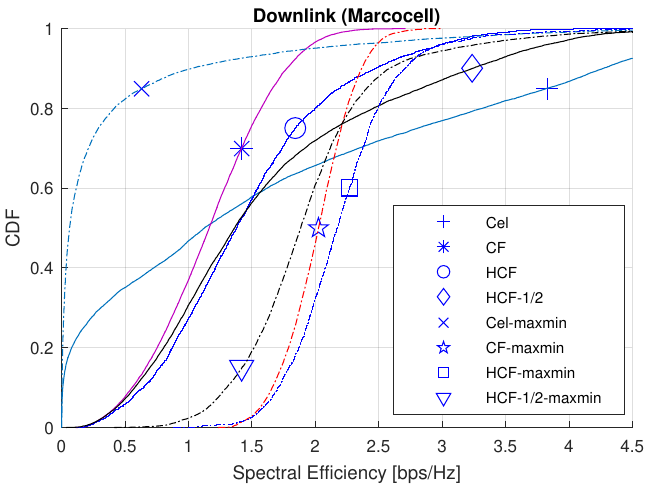}
\label{fig:result2}
}
}
\hspace{15mm}
 \caption{Downlink performance comparison of HCF, CF, and cellular systems in (a) microcell and (b) marcocell under equal and max-min power control.    }
 \label{fig:result}
\end{figure*}

\begin{figure*}[!tbph]
\centerline{
\subfloat[]{
\includegraphics[width=0.4\textwidth]{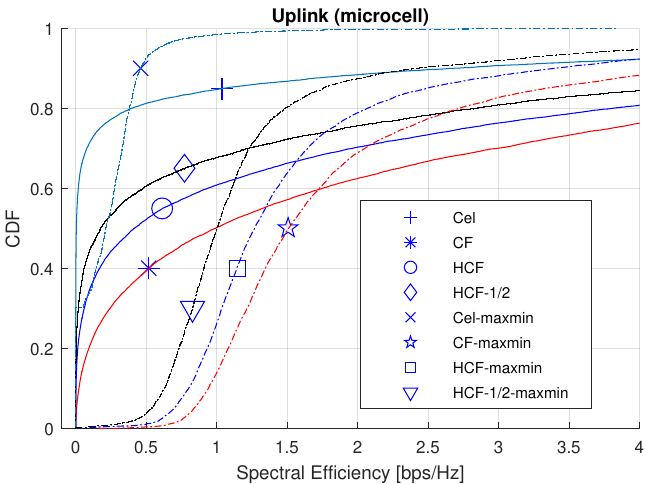}
\label{fig:result3} 
}
\hspace{20mm}
\subfloat[]{
\includegraphics[width=0.4\textwidth]{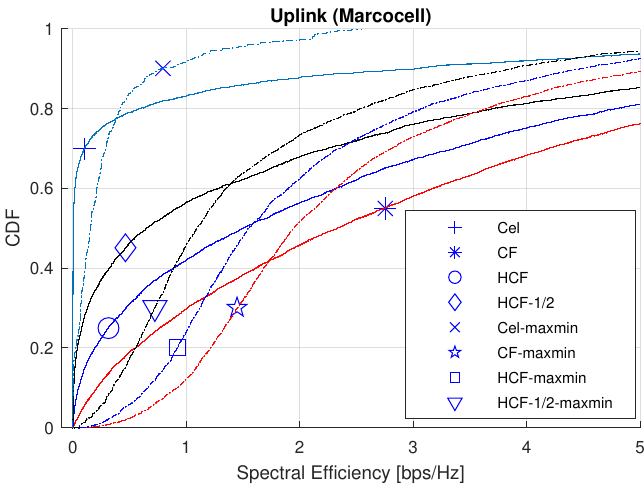}
\label{fig:result4}
}
}
\hspace{15mm}
 \caption{Uplink performance comparison of HCF, CF, and cellular systems in (a) microcell and (b) marcocell under \tcr{full power} and max-min power control.  }
 \label{fig:result2222}
\end{figure*}

\subsection{Numerical Results}
We begin by examining the downlink performance in a microcell scenario, comparing the cumulative distribution functions (CDFs) of per-user SE for HCF, CF, and cellular (cel) networks, under both equal and max-min power control, as illustrated in \figurename \ref{fig:result1}.  The $95\%\text{-likely}$ user rate, as a key indicator for cell-edge performance, corresponds to the $5^{th}$ percentile of these CDF curves. HCF  (default with 1/4 antenna aggregation) achieves a $95\%\text{-likely}$ rate of $0.65\mathrm{bps/Hz}$. This result outperforms CF ($0.61\mathrm{bps/Hz}$), and demonstrates a 50-fold improvement compared to cellular ($0.013\mathrm{bps/Hz}$). Max-min power control significantly boosts performance for distributed architectures: CF-maxmin reaches $2.05\mathrm{bps/Hz}$, while HCF-maxmin achieves $2.03\mathrm{bps/Hz}$. In cellular systems, the $95\%\text{-likely}$ rate degrades under max-min power control ($0.006\mathrm{bps/Hz}$ vs. $0.013\mathrm{bps/Hz}$). This happens because the majority of the total power—sometimes over $90\%$—is assigned to a cell-edge user with the weakest channel, leaving very little power for cell-center users with stronger channels. Such extreme power redistribution is highly inefficient. Power control operates differently in HCF: each distributed eAP independently allocates most of its power to its own worst-connected users without affecting other eAPs. This localized optimization ensures fairness without sacrificing overall performance. The effect of HCF’s antenna aggregation are further examined. Increasing centralization to HCF-1/2 (half of antennas aggregated centrally) reduces the results under equal power ($0.55\mathrm{bps/Hz}$ vs. $0.65\mathrm{bps/Hz}$ for default HCF). However, it reduces fronthaul cost and signaling overhead by roughly $50\%$ (compared to $25\%$ savings for default HCF). Applying max-min to HCF-1/2 recovers significant fairness  ($1.72\mathrm{bps/Hz}$), though it remains below CF-maxmin, illustrating its balance between centralization benefits (cost savings) and distributed gains (fairness).

In our extended evaluation, as shown in \figurename \ref{fig:result2}, we evaluated downlink performance in a macrocell scenario. As expected, user rates declined compared to that of microcell due to larger propagation loss. HCF achieves $0.46~\mathrm{bps/Hz}$, outperforming cellular ($0.0017$) and CF ($0.42~\mathrm{bps/Hz}$), reaffirming its superiority in large-coverage cases. Max-min power control significantly enhances fairness of HCF systems to $1.59~\mathrm{bps/Hz}$, surpassing even CF-maxmin ($1.56~\mathrm{bps/Hz}$). Conversely, cellular systems exhibit performance degradation under max-min (less than $0.001$  vs. $0.0017~\mathrm{bps/Hz}$), as extreme inefficient power allocation to cell-edge users starves cell-center users. For HCF-1/2, $95\%\text{-likely}$ rate declines under equal power ($0.43$ vs. $0.46~\mathrm{bps/Hz}$ for default HCF). Applying max-min to HCF-1/2  partially recovers fairness ($1.15~\mathrm{bps/Hz}$) but remains below CF-maxmin. The proposed max-min power control not only enhances performance but also effectively reduces transmission power. Specifically, it achieves a $\mathbf{10.67\%}$ reduction in power at the cBS and an average reduction of $\mathbf{23.51\%}$ in power usage for eAPs.

The simulation results for uplink microcell scenarios is given in \figurename \ref{fig:result3}. Cellular networks exhibit  poor performance, highlighting the inefficiency of traditional cellular systems for edge users. In contrast, CF and HCF demonstrate vast superiority, particularly with max-min power control, achieving $95\%\text{-likely}$ user rates of 0.83 and 0.70 bps/Hz, respectively. While HCF loses some worst-case performance compared to CF, it still substantially outperforms the cellular network. Importantly, HCF achieves high fairness while reducing fronthaul scale and signaling overhead by approximately $25\%$. HCF-1/2 trades additional performance for even greater cost efficiency, showing further rate loss (with a $95\%\text{-likely}$ rate of 0.54 bps/Hz), but achieving an impressive $50\%$ reduction in fronthaul network, making it a highly cost-effective alternative. 

\begin{figure}[!tbph]
    \centering
    \includegraphics[width=0.4\textwidth]{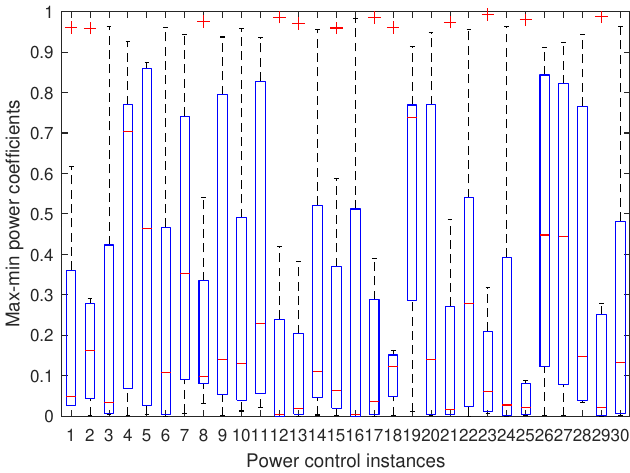}
    \caption{Variations in optimal max-min power coefficients.    }
    \label{fig:power}
\end{figure}

Max-min power control across users is highly effective, not only enhancing performance  but also substantially reducing the transmission power, \tcr{compared with the full power transmission of UEs}. \figurename \ref{fig:power} illustrates the variation in optimal coefficients across 30 independent power control trials. The Y-axis represents the power coefficient, ranging from 0 to 1, while the X-axis corresponds to independent power control instances over different simulation epoch, where user locations are randomly varied. Each box plot features a central line indicating the median, box edges marking the $25^{th}$ and $75^{th}$ percentiles, and a cross symbol representing the peak transmission power among users. The data reveals that while the worst-performing users transmit almost full power, the majority of users operate at lower power levels. Aggregating results from 2000 epochs, the proposed max-min algorithm saves approximately $\mathbf{73}\%$ power, \tcr{compared with simply using full power in the uplink}. This is particularly advantageous for battery-powered mobile terminals, extending their operational lifetime.

%Last but not least, the uplink max-min power control requires optimizing only $K$ coefficients, making it computationally manageable and significantly simpler than the downlink control, which involves at most $K\times M$ coefficients. 

The simulation results for uplink macrocell scenarios, as illustrated in \figurename \ref{fig:result4}, demonstrate performance trends consistent with those observed in microcell environments. 
The proposed hierarchical architecture achieves a $95\%\text{-likely}$ rate of 0.015 bps/Hz under equal power conditions, which is lower than that of CF but still significantly outperforms cellular networks. Increasing centralization to HCF-1/2 leads to a performance decline to below 0.002 bps/Hz. This highlights the important role of distributed APs in promoting fairness. Implementing max-min power control within the HCF and HCF-1/2  enhances performance to 0.45 bps/Hz and 0.23 bps/Hz, respectively, partially restoring the fairness observed in CF (0.68 bps/Hz). Notably, max-min power control reduces the transmission power of user equipment by approximately $\mathbf{60.5\%}$ in this case.

\begin{figure}[!tbph]
    \centering
    \includegraphics[width=0.4\textwidth]{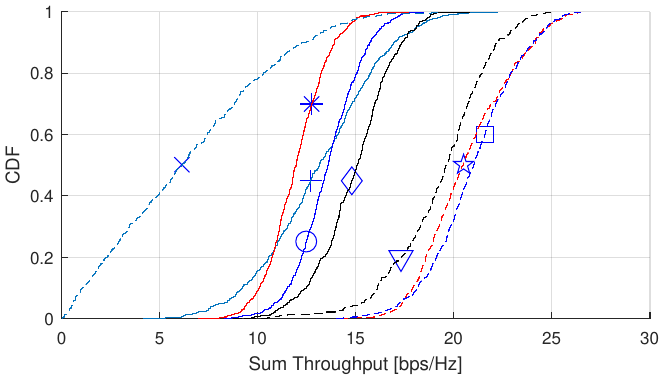}
    \caption{System Capacity Comparison. Note that: marker identifiers align with those in previous figures.    }
    \label{fig:sumThroughput}
\end{figure}

In addition to fairness, overall system capacity serves as a critical performance metric. \figurename \ref{fig:sumThroughput} illustrates the CDFs of sum throughput for HCF, CF, and cellular networks under both equal and max-min power control in the downlink of a microcell scenario, as \figurename \ref{fig:result1}. The results reveal distinct trade-offs: cellular systems, despite their severe cell-edge performance limitations, achieve high sum throughput due to the exceptional QoS experienced by cell-center users. Conversely, fully distributed CF networks prioritize fairness at the expense of lower sum throughput. HCF, particularly HCF-1/2, strike a balance by leveraging cBS to maintain high QoS for cell-center users while deploying eAPs to enhance fairness. Our approach optimizes both fairness and system capacity while minimizing fronthaul infrastructure demands. Under max-min power control, as expected, cellular networks suffer substantial performance degradation due to inefficient power allocation. Although max-min power control improves CF network throughput, HCF consistently outperform CF, demonstrating superior capacity-fairness trade-offs.

%The 5 percentile of per-user Spectral Efficiency for cellular, cellular with max-min power control, CF, CF-Maxmin, HCF, HCF-Maxmin, HCF-1/2, and HCF-1/2_maxmin is 0.012841 bps/Hz, 0.005827 bps/Hz, 0.612224 bps/Hz, 2.051955 bps/Hz, 0.653616 bps/Hz, 2.027435 bps/Hz, 0.550387 bps/Hz, 1.720505 bps/Hz
%The 5 percentile of per-user Spectral Efficiency in Marcocell downlink is  cellular (0.0017 bps/Hz), cellular with max-min power control (9.2692e-04), CF 0.4204, CF-Maxmin 1.5630, HCF 0.4567 bps/Hz, HCF-Maxmin 1.5882, HCF-1/2 0.4297 , and HCF-1/2_maxmin  1.1518
%The 95%-likely per-user Spectral Efficiency in microcell uplink is  cellular (1.5903e-08), cellular with max-min power control (0), CF 0.0063, CF-Maxmin 78, HCF 2.0528e-04 bps/Hz, HCF-Maxmin 0.6974, HCF-1/2 2.4103e-05 , and HCF-1/2_maxmin  0.5396
%The 95%-likely per-user Spectral Efficiency in Macrocell uplink is  cellular (0), cellular with max-min power control (0), CF 0.0905, CF-Maxmin 0.6767, HCF 0.0148 bps/Hz, HCF-Maxmin 0.4366, HCF-1/2 0.0016 , and HCF-1/2_maxmin 0.1661

\section{Conclusion}

Traditional cellular systems suffer from severe performance degradation for cell-edge users, while fully distributed cell-free systems impose high fronthaul complexity and costs. This work introduces a hierarchical cell-free massive MIMO design, which aggregates a subset of antennas at a central base station while retaining some edge access points, thereby reducing fronthaul scale. Numerical results demonstrate that HCF ensures fairness, achieving a $95\%$-likely per-user spectral efficiency comparable to CF while significantly outperforming cellular networks. In addition to offering uniform quality of service, HCF preserves the benefits of centralization—e.g., HCF-1/2 reduces fronthaul scale by $50\%$ while getting much higher system capacity than CF. Furthermore, the proposed max-min algorithms not only enhance fairness and system capacity but also substantially save power—particularly attractive for battery-powered user devices.

\appendices
\section{Expectations under pilot contamination} \label{app_prelim}
The calculations for $\mathbb{E}[|\mathbf{h}_{kl}^T \hat{\mathbf{h}}^*_{k'l} |^2]$  vary depending on whether user $k'$ shares the pilot sequence of user $k$.  For $k'\notin \mathcal{P}_k$, $\mathbf{h}_{kl}$ and $\hat{\mathbf{h}}_{k'l}$ are \textit{independent} because of $\mathbb{E}[\mathbf{i}_k^H \mathbf{i}_{k'}]=0$. Thus, we have
\begin{align} \nonumber \label{eQN:expeResults}
     \mathbb{E}[|\mathbf{h}_{kl}^T \hat{\mathbf{h}}^*_{k'l} |^2] &=   \mathbb{E}[ \mathbf{h}_{kl}^T \hat{\mathbf{h}}^*_{k'l} \hat{\mathbf{h}}^T_{k'l}\mathbf{h}_{kl}^* ] \\  \nonumber
     &= \mathrm{tr} \left ( \mathbb{E}[ \mathbf{h}_{kl}^*\mathbf{h}_{kl}^T  ] \mathbb{E}[ \hat{\mathbf{h}}^*_{k'l} \hat{\mathbf{h}}^T_{k'l} ]  \right) \\ 
     &= p_u \tau_p  \tr \left ( \mathbf{R}_{kl} \mathbf{R}_{k'l} \boldsymbol{\Gamma}_{k'l}^{-1}\mathbf{R}_{k'l}    \right),
\end{align}
due to $\mathbb{E}[ \mathbf{h}_{kl}\mathbf{h}_{kl}^H  ]=\mathbf{R}_{kl}$ and $\mathbb{E}[ \hat{\mathbf{h}}_{k'l} \hat{\mathbf{h}}^H_{k'l} ]=p_u \tau_p\mathbf{R}_{k'l} \boldsymbol{\Gamma}_{k'l}^{-1}\mathbf{R}_{k'l}   $ see \eqref{EQN_correlationMatrixCHest}.
In contrast, $\mathbf{h}_{kl}$ and $\hat{\mathbf{h}}_{k'l}$ are \textit{correlated} for $k'\in \mathcal{P}_k$ (including $k$) due to pilot contamination caused by the shared pilot sequence. Consequently, the mathematical expectation becomes \textit{non-multiplicative}, where \(
\mathbb{E}[X_1X_2] \neq \mathbb{E}[X_1] \mathbb{E}[X_2] \) for two correlated random variables \(X_1\) and $X_2$.
Thus, 
\begin{align}  \label{EQN_correltedEhh}
     \mathbb{E}[|\mathbf{h}_{kl}^T \hat{\mathbf{h}}^*_{k'l} |^2] 
     &=  \mathbb{E}[ (\hat{\mathbf{h}}_{kl}^T+\tilde{\mathbf{h}}_{kl}^T) \hat{\mathbf{h}}^*_{k'l} \hat{\mathbf{h}}^T_{k'l}(\hat{\mathbf{h}}_{kl}^*+\tilde{\mathbf{h}}_{kl}^*) ] \\  \nonumber
     &= \underbrace{ \mathbb{E}[ \hat{\mathbf{h}}_{kl}^T \hat{\mathbf{h}}^*_{k'l} \hat{\mathbf{h}}^T_{k'l}\hat{\mathbf{h}}_{kl}^* ] }_{\mathcal{F}_1} + \underbrace{\mathbb{E}[ \tilde{\mathbf{h}}_{kl}^T \hat{\mathbf{h}}^*_{k'l} \hat{\mathbf{h}}^T_{k'l}\tilde{\mathbf{h}}_{kl}^* ]}_{\mathcal{F}_2}, 
\end{align}
where the two cross terms vanish due to $\mathbb{E}[ \hat{\mathbf{h}}_{kl}^H\tilde{\mathbf{h}}_{kl}]=0 $.

Recalling \eqref{eQn_hhat}, we have $\hat{\mathbf{h}}_{kl} =  \sqrt{p_u} \mathbf{R}_{kl} \boldsymbol{\Gamma}_{kl}^{-1} \boldsymbol{\psi}_{kl}$, and similarly $\hat{\mathbf{h}}_{k'l} =  \sqrt{p_u} \mathbf{R}_{k'l} \boldsymbol{\Gamma}_{kl}^{-1} \boldsymbol{\psi}_{kl}$. Here, $\boldsymbol{\Gamma}_{k'l}=\boldsymbol{\Gamma}_{kl}$ and $\boldsymbol{\psi}_{k'l}=\boldsymbol{\psi}_{kl}$ because of the shared pilot sequence.
Referring to Corollary 4.5 of \cite{SIG-093}, $\mathcal{F}_1$ in \eqref{EQN_correltedEhh} becomes 
\begin{align} \nonumber \label{eQn_FirstIntTerm}
    \mathcal{F}_1=&  \mathbb{E}[ |\hat{\mathbf{h}}^H_{k'l}\hat{\mathbf{h}}_{kl} |^2 ] = p_u^2\mathbb{E}[ | \boldsymbol{\psi}_{kl}^H  \boldsymbol{\Gamma}_{kl}^{-1} \mathbf{R}_{k'l}   \mathbf{R}_{kl} \boldsymbol{\Gamma}_{kl}^{-1} \boldsymbol{\psi}_{kl} |^2 ]\\ 
    = & p_u^2 \tau_p^2\left| \mathrm{tr}( \boldsymbol{\Gamma}_{kl}^{-1} \mathbf{R}_{k'l}   \mathbf{R}_{kl} \boldsymbol{\Gamma}_{kl}^{-1} \boldsymbol{\Gamma}_{kl}  )  \right|^2\\ \nonumber & + p_u^2\tau_p^2\mathrm{tr}(  \boldsymbol{\Gamma}_{kl}^{-1} \mathbf{R}_{k'l}   \mathbf{R}_{kl} \boldsymbol{\Gamma}_{kl}^{-1} \boldsymbol{\Gamma}_{kl} \boldsymbol{\Gamma}_{kl}^{-1}  \mathbf{R}_{kl}  \mathbf{R}_{k'l} \boldsymbol{\Gamma}_{kl}^{-1} \boldsymbol{\Gamma}_{kl} )\\ \nonumber
    = & p_u^2 \tau_p^2 \left| \mathrm{tr}( \boldsymbol{\Gamma}_{kl}^{-1} \mathbf{R}_{k'l}   \mathbf{R}_{kl}   )  \right|^2 \\ \nonumber & 
    + p_u^2 \tau_p^2\mathrm{tr}(  \boldsymbol{\Gamma}_{kl}^{-1} \mathbf{R}_{k'l}   \mathbf{R}_{kl} \boldsymbol{\Gamma}_{kl}^{-1}  \mathbf{R}_{kl}  \mathbf{R}_{k'l}  ) \\ \nonumber
    = & p_u^2 \tau_p^2 \left| \mathrm{tr}(     \mathbf{R}_{kl}  \boldsymbol{\Gamma}_{kl}^{-1}\mathbf{R}_{k'l} )  \right|^2 \\ \nonumber & 
    + p_u\tau_p\mathrm{tr}(    ( \mathbf{R}_{kl} - \boldsymbol{\Theta}_{kl} )  \mathbf{R}_{k'l} \boldsymbol{\Gamma}_{kl}^{-1} \mathbf{R}_{k'l} ).
\end{align}
The above derivation relies on the fact that $\mathbf{R}_{kl}$, for any $k$, is a Hermitian matrix, i.e., $\mathbf{R}_{kl}=\mathbf{R}_{kl}^H$, and the cyclic property of the trace operation, which allows  
\begin{equation}
    \tr(\mathbf{R}_{kl}\boldsymbol{\Gamma}_{kl}^{-1}\mathbf{R}_{k'l}) = \tr(  \boldsymbol{\Gamma}_{kl}^{-1}\mathbf{R}_{k'l}\mathbf{R}_{kl})=\tr(  \mathbf{R}_{k'l}\mathbf{R}_{kl}\boldsymbol{\Gamma}_{kl}^{-1}).
\end{equation} 
The second term $\mathcal{F}_2$ in \eqref{EQN_correltedEhh}  becomes
\begin{align} \label{eQn_secondIntTerm} \nonumber
    \mathcal{F}_2&=\mathbb{E}[ \tilde{\mathbf{h}}_{kl}^H \hat{\mathbf{h}}_{k'l} \hat{\mathbf{h}}^H_{k'l}\tilde{\mathbf{h}}_{kl} ]
    = \mathrm{tr} \left (  \mathbb{E}[\tilde{ \mathbf{h}}_{kl}\tilde{\mathbf{h}}_{kl}^H  ] \mathbb{E}[ \hat{\mathbf{h}}_{k'l} \hat{\mathbf{h}}^H_{k'l} ] \right) \\  
    &= p_u \tau_p \mathrm{tr} \left ( \boldsymbol{\Theta}_{kl} \mathbf{R}_{k'l} \boldsymbol{\Gamma}_{k'l}^{-1}\mathbf{R}_{k'l}    \right).
\end{align}
Substituting \eqref{eQn_FirstIntTerm} and \eqref{eQn_secondIntTerm} into \eqref{EQN_correltedEhh}, yield
\begin{align} \label{eQn:ExpectedValueforhkl}
    \mathbb{E}[|\mathbf{h}_{kl}^T \hat{\mathbf{h}}^*_{k'l} |^2]  = & p_u^2 \tau_p^2 \left| \mathrm{tr}(     \mathbf{R}_{kl}  \boldsymbol{\Gamma}_{k'l}^{-1}\mathbf{R}_{k'l} )  \right|^2  \\ \nonumber
    &   +   p_u \tau_p\mathrm{tr}(   \mathbf{R}_{kl}  \mathbf{R}_{k'l} \boldsymbol{\Gamma}_{k'l}^{-1} \mathbf{R}_{k'l} ) .
\end{align}

This result is also applied for $\mathbf{h}_{k0}$ and $\hat{\mathbf{h}}_{{k'0}}$, we obtain 
\begin{equation} \tcr{
  \label{EQN_preResult_hk0}
  \mathbb{E}[|\mathbf{h}_{k0}^T \hat{\mathbf{h}}_{k'0}^*|^2]=} 
  \left \{\begin{aligned} &p_u \tau_p\tr \left ( \mathbf{R}_{k0} \mathbf{R}_{k'0} \boldsymbol{\Gamma}_{k'0}^{-1}\mathbf{R}_{k'0}    \right), \quad k'\notin \mathcal{P}_k \\
  & \begin{aligned} & p_u^2 \tau_p^2 \left| \mathrm{tr}(     \mathbf{R}_{k0}  \boldsymbol{\Gamma}_{k'0}^{-1}\mathbf{R}_{k'0} )  \right|^2 +\\
    & p_u \tau_p\mathrm{tr}(   \mathbf{R}_{k0}  \mathbf{R}_{k'0} \boldsymbol{\Gamma}_{k'0}^{-1} \mathbf{R}_{k'0} )\end{aligned} , \quad k'\in \mathcal{P}_k
  \end{aligned} \right.
\end{equation}

\section{Proof of Proposition \ref{prop_dl_hcf_se}.} \label{app_proposition_DLHCF}
Referring to \eqref{cfmmimo:formularSNR}, the downlink SINR of HCF is given by
\begin{align}  \label{EQN_DL_sinr_fP} 
    \xi_k^{hcf}  = \frac{|\mathcal{S}_1|^2}{\mathbb{E}\left[|\mathcal{J}_1+\mathcal{J}_2|^2\right] + \sigma_z^2},
\end{align} 
where the power gain of $u_k$ equals
\begin{equation} 
    |\mathcal{S}_1|^2  =    \left |   \sqrt{ p_b\eta_{k0}\mathbb{E}[\|\hat{\mathbf{h}}_{k0}\|^2]}  + \sum\nolimits_{l\in\mathbb{L}}    \sqrt{ p_a \eta_{kl} \mathbb{E}[\|\hat{\mathbf{h}}_{kl}\|^2]}   \right |^2. 
\end{equation}
The expected power of $\hat{\mathbf{h}}_{kl}$ is 
\begin{equation} \label{eQN_expectedhvalued}
    \mathbb{E}[\|\hat{\mathbf{h}}_{kl}\|^2]= \tr\left( \mathbb{E}\left[ \hat{ \mathbf{h}}_{kl} \hat{\mathbf{h}}_{kl}^H \right] \right) = p_u\tau_p \mathrm{tr}\left(\mathbf{R}_{kl} \boldsymbol{\Gamma}_{kl}^{-1}\mathbf{R}_{kl}\right),
\end{equation}
following from \eqref{EQN_correlationMatrixCHest}. Similarly, for $\hat{\mathbf{h}}_{k0}$, we have
\begin{equation} \label{eQN_averageCHpowerGainkl}
    \mathbb{E}[\|\hat{\mathbf{h}}_{k0}\|^2]= p_u\tau_p \mathrm{tr}\left(\mathbf{R}_{k0} \boldsymbol{\Gamma}_{k0}^{-1}\mathbf{R}_{k0}\right),
\end{equation}  since $\mathbb{E}[ \hat{ \mathbf{h}}_{k0} \hat{\mathbf{h}}_{k0}^H ]= p_u\tau_p \mathbf{R}_{k0} \boldsymbol{\Gamma}_{k0}^{-1}\mathbf{R}_{k0}$, as derived from \eqref{eQN_hhatdistribution}.
Thus, 
\begin{align} \label{GS_S1}
    |\mathcal{S}_1|^2  & = \left| \begin{aligned}        
    &\sqrt{p_b\eta_{k0} p_u \tau_p\mathrm{tr}\left(\mathbf{R}_{k0} \boldsymbol{\Gamma}_{k0}^{-1}\mathbf{R}_{k0}\right)}\\  & + \sum \nolimits_{l\in\mathbb{L}} \sqrt{p_a\eta_{kl}p_u \tau_p\mathrm{tr}\left(\mathbf{R}_{kl} \boldsymbol{\Gamma}_{kl}^{-1}\mathbf{R}_{kl}\right)} \end{aligned} \right|^2.
\end{align}

Due to the independence of data symbols, i.e., $\mathbb{E}[u_{k}^*u_{k'}]=0$ for $k'\neq k$, $\mathcal{J}_1$ and $\mathcal{J}_2$ in \eqref{eQn_DLGeneralSig} are uncorrelated. This implies that $\mathbb{E}\left[|\mathcal{J}_1 + \mathcal{J}_2|^2\right]=\mathbb{E}\left[|\mathcal{J}_1|^2\right]+\mathbb{E}\left[|\mathcal{J}_2|^2\right]$. We now compute the variance of $\mathcal{J}_1$ as 
\begin{align}  \label{APPEQ1}
    \mathbb{E}\left[|\mathcal{J}_1|^2\right]  = &   \frac{p_b\eta_{k0} \overbrace{\mathbb{E}\left[ \left| \mathbf{h}_{k0}^T \hat{\mathbf{h}}_{k0}^* - \mathbb{E}[\|\hat{\mathbf{h}}_{k0}\|^2] \right|^2 \right] }^{\mathcal{T}_1} } {\mathbb{E}[\|\hat{\mathbf{h}}_{k0}\|^2]}   \\ \nonumber  + & p_a \sum\nolimits_{l \in \mathbb{L}}  \frac{ \eta_{kl} \overbrace{\mathbb{E}\left[ \left| \mathbf{h}_{kl}^T \hat{\mathbf{h}}_{kl}^* - \mathbb{E}[\|\hat{\mathbf{h}}_{kl}\|^2] \right|^2 \right]}^{\mathcal{T}_2} } {  \mathbb{E}[\|\hat{\mathbf{h}}_{kl}\|^2] } .
\end{align}
The first term $\mathcal{T}_1$ in \eqref{APPEQ1} is further derived as
\begin{align} \label{GS_T1inDL}
    \mathcal{T}_1 = & \mathbb{E}\left[ \left| \mathbf{h}_{k0}^T \hat{\mathbf{h}}_{k0}^*\right|^2 \right] - \mathbb{E}\left[ \mathbf{h}_{k0}^T \hat{\mathbf{h}}_{k0}^* \right]\mathbb{E}[\| \hat{\mathbf{h}}_{k0}\|^2 ]\\ \nonumber 
    & - \mathbb{E}\left[ \mathbf{h}_{k0}^H \hat{\mathbf{h}}_{k0} \right]\mathbb{E}[\| \hat{\mathbf{h}}_{k0}\|^2 ] + \left(\mathbb{E}[\| \hat{\mathbf{h}}_{k0}\|^2 ]\right)^2 \\ \nonumber 
    = & \mathbb{E}\left[ \left| \mathbf{h}_{k0}^T \hat{\mathbf{h}}_{k0}^*\right|^2 \right] - \mathbb{E}\left[ (\hat{\mathbf{h}}_{k0}^T + \tilde{\mathbf{h}}_{k0}^T) \hat{\mathbf{h}}_{k0}^* \right]\mathbb{E}[\| \hat{\mathbf{h}}_{k0}\|^2 ] \\ \nonumber 
    &- \mathbb{E}\left[ (\hat{\mathbf{h}}_{k0}^H + \tilde{\mathbf{h}}_{k0}^H)\hat{\mathbf{h}}_{k0} \right]\mathbb{E}[\| \hat{\mathbf{h}}_{k0}\|^2 ] + \left(\mathbb{E}[\| \hat{\mathbf{h}}_{k0}\|^2 ]\right)^2\\ \nonumber 
    \overset{(a)}{=} & \mathbb{E}\left[ \left| \mathbf{h}_{k0}^T \hat{\mathbf{h}}_{k0}^*\right|^2 \right] - \left(\mathbb{E}[\| \hat{\mathbf{h}}_{k0}\|^2 ] \right)^2.
\end{align}
The derivation in step $(a)$ relies on the independence between a channel estimate and its estimation error, namely $\mathbb{E}[ \tilde{\mathbf{h}}_{k0}^H\hat{\mathbf{h}}_{k0} ]=0 $.
From \eqref{EQN_preResult_hk0} in Appendix \eqref{app_prelim}, we have 
\begin{align} 
  \label{GS_T1_firstterm}  \nonumber
  \mathbb{E}[|\mathbf{h}_{k0}^T \hat{\mathbf{h}}_{{k0}}|^2]= & 
 p_u^2 \tau_p^2 \left( \mathrm{tr}(     \mathbf{R}_{k0}  \boldsymbol{\Gamma}_{k0}^{-1}\mathbf{R}_{k0} )  \right)^2 \\ &+ p_u \tau_p\mathrm{tr}(   \mathbf{R}_{k0}  \mathbf{R}_{k0} \boldsymbol{\Gamma}_{k0}^{-1} \mathbf{R}_{k0} ).
\end{align}
Substituting \eqref{eQN_averageCHpowerGainkl} and \eqref{GS_T1_firstterm} into \eqref{GS_T1inDL}, $\mathcal{T}_1$ becomes
\begin{equation} 
    \mathcal{T}_1 = p_u \tau_p\mathrm{tr}(   \mathbf{R}_{k0}  \mathbf{R}_{k0} \boldsymbol{\Gamma}_{k0}^{-1} \mathbf{R}_{k0} ).
\end{equation}
Similarly,
\begin{equation}
    \mathcal{T}_2 = p_u \tau_p\mathrm{tr}(   \mathbf{R}_{kl}  \mathbf{R}_{kl} \boldsymbol{\Gamma}_{kl}^{-1} \mathbf{R}_{kl} ).
\end{equation}
Using \eqref{eQN_expectedhvalued}, \eqref{eQN_averageCHpowerGainkl}, along with the expressions of $\mathcal{T}_1$ and $\mathcal{T}_2$, the following result is obtained:
\begin{align} \label{GSVarianceIone}
    \mathbb{E}\left[|\mathcal{J}_1|^2\right]= & \frac{p_b\eta_{k0} \mathrm{tr}(   \mathbf{R}_{k0}  \mathbf{R}_{k0} \boldsymbol{\Gamma}_{k0}^{-1} \mathbf{R}_{k0} )} {\mathrm{tr}\left(\mathbf{R}_{k0} \boldsymbol{\Gamma}_{k0}^{-1}\mathbf{R}_{k0}\right)   }   \\ \nonumber & +  p_a \sum\nolimits_{l \in \mathbb{L}}  \frac{\eta_{kl} \mathrm{tr}(   \mathbf{R}_{kl}  \mathbf{R}_{kl} \boldsymbol{\Gamma}_{kl}^{-1} \mathbf{R}_{kl} )} {\mathrm{tr}\left(\mathbf{R}_{kl} \boldsymbol{\Gamma}_{kl}^{-1}\mathbf{R}_{kl}\right)   }. 
\end{align}

Next, the variance of $\mathcal{J}_2$ is computed as
\begin{align} \label{GS_VarianceJ2} \nonumber
    \mathbb{E}\left[|\mathcal{J}_2|^2\right]= &  p_b\sum\nolimits_{k'\in \mathbb{K} \backslash \{k\}}  \frac{ \eta_{k'0} \mathbb{E}[|\mathbf{h}_{k0}^T \hat{\mathbf{h}}^*_{k'0}|^2] }{  \mathbb{E}[\|\hat{\mathbf{h}}_{k'0}\|^2]  } \\  
     + & p_a  \sum\nolimits_{k'\in \mathbb{K} \backslash \{k\} } \sum\nolimits_{l\in\mathbb{L}}  \frac{ \eta_{k'l} \mathbb{E}[|\mathbf{h}_{kl}^T\hat{\mathbf{h}}^*_{k'l}|^2] } { \mathbb{E}[\|\hat{\mathbf{h}}_{k'l}\|^2] }  . 
\end{align}
Based on the presence of pilot contamination, the set $\mathbb{K} \backslash \{k\}$ is divided into two non-overlapped subsets: $ \mathcal{P}_k \backslash \{k\}$ and $ \mathbb{K}\backslash\mathcal{P}_k $.
Applying \eqref{eQN:expeResults}, \eqref{eQn:ExpectedValueforhkl}, \eqref{EQN_preResult_hk0}, \eqref{eQN_expectedhvalued}, and \eqref{eQN_averageCHpowerGainkl}, the variance of $\mathcal{J}_2$ is obtained as 
\begin{align} \label{GS_VarianceJ2} \nonumber
    \mathbb{E}&\left[|\mathcal{J}_2|^2\right]= \\ \nonumber &  p_b\sum\nolimits_{k'\in \mathcal{P}_k \backslash \{k\}}  \frac{ \eta_{k'0} \left\{ \begin{aligned} &p_u \tau_p \left| \mathrm{tr}(     \mathbf{R}_{k0}  \boldsymbol{\Gamma}_{k'0}^{-1}\mathbf{R}_{k'0} )  \right|^2 \\
    & +\mathrm{tr}(   \mathbf{R}_{k0}  \mathbf{R}_{k'0} \boldsymbol{\Gamma}_{k'0}^{-1} \mathbf{R}_{k'0} )\end{aligned}\right\} }{   \mathrm{tr}\left(\mathbf{R}_{k'0} \boldsymbol{\Gamma}_{k'0}^{-1}\mathbf{R}_{k'0}\right)  } \\ \nonumber 
    &+  p_b\sum\nolimits_{k'\in \mathbb{K} \backslash \mathcal{P}_k }  \frac{ \eta_{k'0}   \tr \left ( \mathbf{R}_{k0} \mathbf{R}_{k'0} \boldsymbol{\Gamma}_{k'0}^{-1}\mathbf{R}_{k'0}    \right) }{   \mathrm{tr}\left(\mathbf{R}_{k'0} \boldsymbol{\Gamma}_{k'0}^{-1}\mathbf{R}_{k'0}\right)  } \\
     &+  p_a  \sum\nolimits_{k'\in \mathcal{P}_k \backslash \{k\}} \sum\limits_{l\in\mathbb{L}}  \frac{ \eta_{k'l} \left\{ \begin{aligned} &p_u \tau_p \left| \mathrm{tr}(     \mathbf{R}_{kl}  \boldsymbol{\Gamma}_{k'l}^{-1}\mathbf{R}_{k'l} )  \right|^2 \\ \nonumber
    & +\mathrm{tr}(   \mathbf{R}_{kl}  \mathbf{R}_{k'l} \boldsymbol{\Gamma}_{k'l}^{-1} \mathbf{R}_{k'l} )\end{aligned}\right\} }{   \mathrm{tr}\left(\mathbf{R}_{k'l} \boldsymbol{\Gamma}_{k'l}^{-1}\mathbf{R}_{k'l}\right)  } \\  
    &+  p_a\sum\nolimits_{k'\in \mathbb{K} \backslash \mathcal{P}_k }  \frac{ \eta_{k'l}   \tr \left ( \mathbf{R}_{kl} \mathbf{R}_{k'l} \boldsymbol{\Gamma}_{k'l}^{-1}\mathbf{R}_{k'l}    \right) }{   \mathrm{tr}\left(\mathbf{R}_{k'l} \boldsymbol{\Gamma}_{k'l}^{-1}\mathbf{R}_{k'l}\right)  }  
\end{align}

In final, substituting \eqref{GS_S1}, \eqref{GSVarianceIone}, and \eqref{GS_VarianceJ2} into \eqref{EQN_DL_sinr_fP}, and rearranging the terms yields \eqref{eQN_SINR_DL_MR_final}. The final derivation applies 
\[\frac{\left| \mathrm{tr}(     \mathbf{R}_{kl}  \boldsymbol{\Gamma}_{kl}^{-1}\mathbf{R}_{kl} )  \right|^2}{ \mathrm{tr}(     \mathbf{R}_{kl}  \boldsymbol{\Gamma}_{kl}^{-1}\mathbf{R}_{kl} )}= \left| \mathrm{tr}(     \mathbf{R}_{kl}  \boldsymbol{\Gamma}_{kl}^{-1}\mathbf{R}_{kl} )  \right|\] given that $\mathrm{tr}(     \mathbf{R}_{kl}  \boldsymbol{\Gamma}_{kl}^{-1}\mathbf{R}_{kl} )= \frac{
    \mathbb{E}[\|\hat{\mathbf{h}}_{kl}\|^2]}{p_u\tau_p} > 0$, recalling \eqref{eQN_expectedhvalued}.
\bibliographystyle{IEEEtran}
\bibliography{IEEEabrv,Ref_TCOM}

\end{document}